\documentclass[12pt,preprint]{emulateapj}
\newcommand{\degree}{\ensuremath{^\circ}}
\begin{document}

\title{Tests of the Radial Tremaine-Weinberg Method}
\author{Sharon E. Meidt and Richard J. Rand}
\affil{Department of Physics and Astronomy, \\University of New Mexico, 800 Yale Blvd Northeast, Albuquerque, NM 87131}
\author{Michael R. Merrifield}
\affil{University of Nottingham, School of Physics $\&$ Astronomy, University Park, Nottingham, NG7 2RD}
\author{Victor P. Debattista}
\affil{Centre For Astrophysics, University of Central Lancashire, Preston, UK PR1
2HE}\and
\author{Juntai Shen}
\affil{McDonald Observatory, The University of Texas at Austin, 1 University
Station, C1402, Austin, TX 78712}

\begin{abstract}
At the intersection of galactic dynamics, evolution and global structure, issues such as the relation between bars and spirals 
and the persistence of spiral patterns can be addressed through the characterization of the angular speeds of the patterns and their possible radial variation.  The Radial Tremaine-Weinberg (TWR) Method, a generalized version 
of the Tremaine-Weinberg method for observationally determining a single, constant pattern speed, allows the pattern 
speed to vary arbitrarily with radius.  Here, we perform tests of the 
TWR method with regularization on several simulated galaxy data sets.  The regularization is employed as a means of smoothing intrinsically noisy solutions, as well as for testing model solutions of different radial dependence (e.g. constant, linear or quadratic).  We test these facilities in studies of individual simulations, and demonstrate successful measurement of both bar and spiral pattern speeds in a single disk, secondary bar pattern speeds, and spiral winding (in the first application of a TW calculation to a spiral simulation).  We also explore the major sources of error in the calculation and find uncertainty in the major axis position angle most dominant.  In all cases, the method is able to extract pattern speed solutions where discernible patterns exist to within $20\%$ of the known values, suggesting that the TWR method should be a valuable tool in the area of galactic dynamics.  For utility, we also discuss the caveats in, and compile a prescription for, applications to real galaxies.
\end{abstract}

\section{Introduction}
One of the prime unresolved issues in the 
dynamics and evolution of galaxy disks remains the origin and evolution of large-scale bar and spiral structure.  Though the 
persistence of grand-design
spirals has been tied observationally to the presence of bars or companions \citep{kn79}, virtually 
nothing is known about the actual lifetimes of spiral patterns.  Additionally, despite indications that the relation 
between bar and spiral pattern speeds (which \citet{ss88} first argued may not be equal) 
may be important 
for understanding the role of bars in angular momentum transfer during secular disk evolution (e.g. \citealt{ds98} and \citealt{ds00}), 
there are as yet unanswered questions about the connections between multiple patterns in different radial zones.  While 
mode-coupling between patterns, which allows efficient outward angular momentum transfer in disks (\citealt{syg}; 
\citealt{mt97}) seems a most promising link, in 2D N-body simulations with a dissipative gas component \citet{rs} find evidence for spiral 
structure in the absence of a bar, bar-spiral mode coupling, spiral-spiral mode coupling, and multiple pattern speeds 
without mode coupling.

Clearly, to address questions about the persistence of spiral patterns and the 
relation between bars and spirals requires not only determination of the pattern speed but how it varies with 
radius; only with accurate measurement of bar and spiral or inner and outer spirals pattern speeds in the 
same galaxy can we confirm whether spiral structure is steady or winding, whether bars and spiral pattern speeds are 
equal or are unrelated, whether mode-coupling exists, and the domain and number of patterns that can be sustained in 
a disk.

Because they are not directly accessible through observation, pattern speeds are often determined with indirect means such as the identification of predicted behavior at resonance radii (e.g. \citealt{elm89}; \citealt{elm96}) or kinematic and morphological comparisons of simulated and observed structure (e.g. \citealt{raut2005}; \citealt{gb}).  It is also clearly desirable to employ methods for estimating pattern speeds which do not rely on theoretical models or simulation.  Many other pattern speed determinations have therefore centered on the use of the model-independent method of Tremaine \& Weinberg (1984; hereafter TW) which presents a rigorous derivation for the pattern speed $\Omega_p$ based on the requirement of continuity using observationally accessible quantities.  The determination of $\Omega_p$ involves 
surface density-weighted position and 
velocity line integrals parallel to the galaxy major axis under several essential assumptions.  Specifically, 
the method requires that
the disk of the galaxy is flat (unwarped) and contains a single, well-defined rigidly-rotating pattern; that the 
surface density of a kinematic tracer of a disk component, which must obey continuity, becomes negligibly small at some radius and 
all azimuths within the map boundary (thereby critically yielding converged integrals); and that the relation between the emission from this component and its surface density is
linear, or if not, suspected deviations from linearity can be modeled.

In that information from all sampled radii is associated with a single, constant pattern speed, the TW calculation poses a challenge for 
extracting multiple distinct or radially varying pattern speeds. Non-axisymmetric structures beyond a dominant pattern such as a bar in the disk of a galaxy will interfere with the 
measurement of $\Omega_p$ of the bar; when a non-axisymmetric disk can be decomposed into two components with different
pattern speeds, then the TW estimate is a luminosity- and asymmetry-weighted average of the two patterns \citep{dgs02}.  The TW estimate for the secondary pattern speed in the inner disk of NGC 6946 observed in H$_\alpha$, for example,  is estimated to be limited to an uncertainty of as much as 50\% given the primary pattern's contribution to the TW integrals \citep{fathi}.

Such issues notwithstanding, there have been several recent adaptations of the TW 
method to positive effect.  Applications of the TW method to SB0 galaxies using stellar light as a 
tracer extend the limits of integration in the TW calculation to just past the end of the bar in order to minimize 
contributions to the TW integrals from non-axisymmetric (and several magnitudes dimmer) features beyond the bar; in such cases, integrating past the structure of interest is found to be sufficient for achieving converged integrals.  The 
bar pattern speeds in NGC 7079 \citep{dw04} and NGC 1023 \citep{dca02}, for example, have both been successfully measured in this way.

When there exist more than one pattern in distinct radial zones, however, arguments about the 
convergence of the TW integrals are less straight-forward.  To measure the secondary bar pattern speed in NGC 2950, \citet{cda03}
and \citet{mac06} explore decoupling the inner secondary and outer primary bar pattern speeds 
by associating each component with unique surface brightness contributions in the TW calculation.  The TW integrals are 
modified by the presence of the inner pattern based on an assumption about how and where the two patterns 
decouple (to first order).  This analysis has confirmed the existence of, if not measured, a unique secondary bar pattern speed possibly indicating counter-rotation with respect to the primary bar.

At their best--improving the accuracy of pattern speed estimates to about 20\% \citep{GD}--these kinds of 
adaptations of the TW method for measurement of single or multiple patterns still require assumptions about bar extent 
based on morphological or kinematic signatures.  
To separate the observed surface brightness in NGC 2950 into secondary and primary bars components, \citet{mac06} must 
assume that the patterns indeed decouple at the inner-bar end, or 
that the outer pattern is axially symmetric at least within the inner's extent.  There, the transition between the two is 
inferred from the location of a plateau in the TW surface brightness-weighted position integral as the limits of 
integration are extended from zero. Perhaps more critically, as investigated in 
$\S$ \ref{sec:SIM1}, the direct association of bar length measured in this manner (or perhaps others) with 
pattern extent may introduce error into TW calculations.

Furthermore, adaptations of the TW method based 
on this type of identification are likely to be inapplicable for spiral pattern estimation.  In spiral galaxies, not 
only can identifying transitions between patterns be less clear, but non-axisymmetric motions are significantly 
smaller (e.g. \citealt{rs87}) than in bars (at least those typically analyzed with the TW method).  Indeed, the TW method has yet to be tested on a simulated spiral.

A recent modification to the TW calculation (Merrifield, Rand \& Meidt 2006; hereafter MRM) in which $\Omega_p$ is allowed 
radial variation promises to be an invaluable resource for tests of long-lived density wave theories and for 
understanding the connection, if any, between bar and spiral pattern speeds.  
Like the TW method, the so-called radial TW (TWR) method uses measurements of observables to extract radially varying 
pattern speeds. As first applied using the BIMA Survey 
of Nearby Galaxies CO observations of the grand-design Sb galaxy NGC 1068 (MRM), the TWR method returned 
a spiral pattern speed solution that declines with radius, allowing a winding time for the pattern to be estimated (e.g. MRM).

As described in MRM, the nature of the discretized calculation presents numerical 
solutions that are highly susceptible to fluctuations as a result of compounded noise in the data.  In this paper 
($\S$ \ref{sec:Regularization}), we develop the TWR method with regularization as a 
means of smoothing intrinsically noisy solutions as well as testing model solutions of different radial dependence (described in $\S$ \ref{sec:Smooth}).  These and other 
commendations notwithstanding, with regularization 
one may 
risk introducing an unrealistic prejudice to TWR solutions.  In $\S$ \ref{sec:Bias}, we address a means of identifying 
when this is likely to occur, and in $\S$ \ref{sec:Fourier} describe a scheme for minimizing such regularization-induced 
bias.

Using evidence that arises from these considerations, as well as other a priori information, theoretically and observationally motivated models 
for $\Omega_p(r)$ can be developed which constrain the number and 
extent of patterns present in a disk.  Once solutions for these models are calculated, the goodness of each must be assessed 
in order to identify the best-fit solution.  In $\S$ \ref{sec:weight}, we outline the criteria with which the models 
are judged and describe our concept of error evaluation in the final solutions.

Beginning in $\S$ \ref{sec:Nbody}, we analyze three simulated galaxies with known pattern speeds (a barred spiral, a slowly winding spiral, and a double barred spiral, marking the first application of the TW method to simulated spiral patterns) in order to develop a general stratagem for application to real galaxies.  As applied to these simulations, we find that the TWR method is able to extract multiple pattern speeds with accuracies on the order of 
(and, as we will see in some cases, better than) the traditional TW method.  With the barred spiral simulation in $\S$ \ref{sec:SIM1} we show that TWR bar pattern speed measurement
presents an improvement over traditional TW bar estimates, particularly when there is evidence of a significant 
contribution from the spiral pattern to the TW integrals.  We find that the regularized TWR method can recover information from both patterns effectively by identifying and treating the bar-to-spiral transition radius (which the TW values 
themselves may not indicate) as a free parameter in the calculation.  In $\S$ \ref{sec:iError}, we analyze the results of the method in detail, particularly with regard to morphological limitations.  We compare 
our TWR results with TW estimates in $\S$ \ref{sec:TWTWR} and examine the influence of systematic errors due to the assumed disk position angle and inclination 
(shown to be crucial for TW estimates in \citet{debPA}) on both in $\S$ \ref{sec:Err}.  We also explore the reliability of a fixed parameterization for the bar-to-spiral transition radius. ($\S$ \ref{sec:Trans}).

In the last third of the paper, we investigate the prospects for extracting spiral pattern speed 
solutions that are winding in nature ($\S$ \ref{sec:SIM2}), marking the first application of the TW method to a spiral simulation, and in $\S$ \ref{sec:SIM3} we address the use of the TWR method 
for the purposes of parameterizing an independently rotating nuclear bar in a double barred simulation. There, the techniques we employ for decoupling and extracting 
measurements of the pattern speeds of both the primary and secondary bar components may present an interesting corollary to recent attempts with the TW method to 
measure secondary bar pattern speeds in the presence of a strong primary bar pattern.  We note here that the TWR method 
is a generalized version of the procedure used on NGC 2950 (see \citealt{cda03, mac06}); separating the surface brightness into two components can 
be thought of as the coarsest version of the discretization that is the back-bone of numerical TWR solutions.

Based on our experience with these simulations, we conclude with comments on the applicability of the 
method to observations of real galaxies in $\S$ \ref{sec:caveats} where we also outline a general prescription for using the 
TWR calculation with regularization.
\section{\label{sec:TWRreg}The TWR Method with Regularization}
\subsection{\label{sec:TWRmethod}The Radial Tremaine-Weinberg Method}
By proceeding under the aforementioned assumptions of Tremaine $\&$ Weinberg (1984), but allowing that $\Omega_p$ may possess spatial variation in 
the radial direction whereby the surface density of the chosen tracer can be written
$\Sigma (x,y,t)=\Sigma(r,\phi-\Omega_p(r)t)$, with appropriate mathematical generalizations, the derivation and 
measurement of the pattern speed $\Omega_p(r)$ can be made from observable intensities and kinematics of a chosen tracer.

Following the derivation given in MRM,
integrating the continuity equation obeyed by the tracer (with the replacement $\partial \Sigma / \partial t=-\Omega_p(r)\partial \Sigma / \partial \phi$)
\begin{equation}
-\Omega_p(r)\frac{\partial{\Sigma}}{\partial{\phi}}+\frac{\partial\Sigma v_x}{\partial x}+\frac{\partial \Sigma v_y}
{\partial y}=0
\label{eq:cont}
\end{equation}
over $x$ and $y$ (thereby eliminating the unobservable $v_x$ and the spatial derivative $\partial / \partial y$), and 
changing from Cartesian to polar coordinates yields
\begin{equation}
\int_{r=y}^\infty \int_{\phi=arcsin(y/r)}^{\pi-arcsin(y/r)} \Omega_p(r)\frac{\partial \Sigma}{\partial \phi}r dr 
d\phi+\int_{-\infty}^\infty \Sigma v_y dx = 0.
\end{equation}
A final integration with respect to $\phi$ results in a Volterra integral equation of the first kind for $\Omega_p(r)$:
\begin{equation}
\int_{r=y}^\infty \left\{[\Sigma(x',y)-\Sigma(-x',y)]r\right\}\Omega_p(r)dr=\int_{-\infty}^\infty \Sigma v_y dx
\label{eq:volt}
\end{equation}
where $x'(r,y)=\sqrt{r^2-y^2}$.

Note that with constant $\Omega_p$ in equation (\ref{eq:volt}), we arrive at the regular TW result
\begin{equation}
\Omega_p \int_{-\infty}^{\infty} \Sigma x dx=\int_{-\infty}^{\infty} \Sigma v_y dx ,
\label{eq:twint}
\end{equation}
which, with normalization by $\int \Sigma dx$ (e.g. \citealt{mk95}), leads to
\begin{equation}
\Omega_p=\frac{<v>}{<x>}
\label{eq:tw}
\end{equation}
where $<$$v$$>$=$\int \Sigma v_y dx / \int \Sigma dx$ and  
$<$$x$$>$=$\int \Sigma x dx / \int \Sigma dx$.

\begin{figure}
\epsscale{.90}
  \plotone{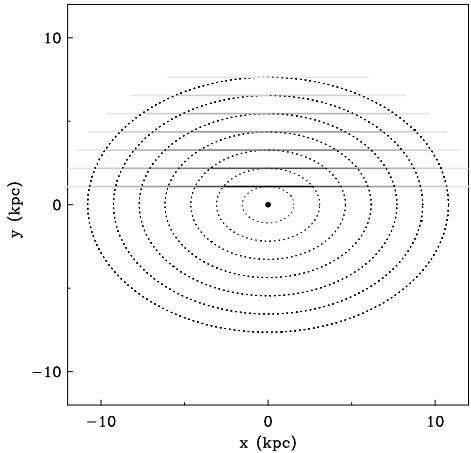}
    \caption{Illustration of a $y$$>$0 quadrature for a galaxy viewed from the negative $z$ side with a tilt around the $y$ axis by $\alpha$=45$\degree$. The horizontal lines, or slices, at positions $y_i$ are spaced at $\Delta y$=1.54, or $\Delta y_{obs}$=1.09, and represent integration between the limits $\pm\sqrt{R_0^2+y_i^2}$ where $R_0$=10.8 is the maximum radial extent of the quadrature.  Each slice is carved into elements of width $\Delta r$ whereby all the elements with the same shade of grey represent a single radial bin $r_j$. \\
\label{fig-quad}}
\end{figure}

For a galaxy projected onto the sky plane with inclination $\alpha$ (so to distinguish from index $i$), both the kernel on the left and the integral on the right 
of equation (\ref{eq:volt}) are observationally determined quantities with $x=x_{obs}$, 
$y=y_{obs}/\cos \alpha$, and $v_y=v_{obs}/\sin \alpha$, where $x_{obs}$ and $y_{obs}$ are the coordinates in the plane of the sky along the major and minor axes, respectively, and $v_{obs}$ is the observed line-of-sight velocity.  Solutions can be extracted numerically by replacing the integral on the 
left with a discrete quadrature for different values of $y=y_i$ and $r=r_j$ (see Figure \ref{fig-quad}) whereby equation (\ref{eq:volt}) is converted to 
\begin{equation}
\Sigma_{r_j>y_i} K(y_i,r_j)\Omega_p(r_j)=b(y_i) 
\end{equation}
or to a matrix equation of the form:
\begin{equation}
 K_{ij}\Omega_j=b_i
\label{eq:twr}
\end{equation}
with $\textbf{K}$ an upper triangular $N\times N$ square matrix.  Note that numerical quadratures on either side of the galaxy ($y$ $<$0 or 
or $y$ $>$0) occur independently, providing two measures of $\Omega_p(r)$.  Furthermore, as governed by the information 
available, the slices which delimit the quadrature on a single side need not be uniformly spaced.  In this case, 
solutions inherit a variable bin width $\Delta r$.  Also, the calculation allows for no azimuthal 
dependence for the pattern speed, which we assume throughout.

The size of $\textbf{K}$ depends on the desired coarseness or fineness of the quadrature; the separation between slices at 
positions $y_i$ (limited by either the resolution or the sampling of the data) translates into a radial bin width (modulo $\cos \alpha$) via equation (\ref{eq:twr}).  
The quadrature, perhaps more critically, depends on the limits
of integration in equation (\ref{eq:volt}).  These limits $\pm X_{max}$ should be chosen
based on where the integrals have converged.  While in 
the case of a single bar pattern integrating past the structure of interest is often suitable, as shown in 
\citet{zrm04} (Figures 9-11), in the presence of strong, extended 
asymmetry TW values are highly dependent on the extent of integration along each slice $i$.  
In cases where multiple patterns exist in a single disk, then, it is equally favorable (and hopefully sufficient) to extend all 
integrals to the edge of the surface brightness distribution.

Meeting the requirement 
of integral convergence in this manner as applied to the TWR calculation determines the location of the last radial bin $j_{max}$
associated with elements $K_{ij_{max}}$ along each slice.  For a given radial bin width, with the requirement that $j_{max}$ equals 
$N$ we are presented with the size of $\textbf{K}$ as well as the outermost slice position, since $j_{max}$ must also equal 
$i_{max}$.  
One should 
check to see that $K_{NN}$, the last entry in $\textbf{K}$ associated with the outermost slice, 
is associated with a fully converged $\int \Sigma x dx$ (which achieves convergence at least by the map boundary).

Since $\textbf{K}$ is an upper triangular matrix, the $\Omega_j$ can be solved for via simple back-substitution. 
In this way, 
solutions are generated from the outermost to the innermost radius (from light gray to black in Figure \ref{fig-quad}) according to 
\begin{equation}
\Omega_k=\frac{b_i- \sum_{j>k}^{N} K_{ij}\Omega_{j}}{K_{kk}}
\label{eq:outin}
\end{equation}
with $k \geq i$ (from eq. \ref{eq:twr}). Such solutions, however, are especially susceptible to wildly oscillatory 
behavior as errors from large radii propagate inward.  These compounded errors arise from uncertainties in the velocities and intensities (which get translated into the 
$K_{ij}$) and can be particularly severe since the outermost bins often cover the lowest signal-to-noise regions in the data.  The errors thus introduced can be systematic.

\begin{figure*}
\begin{center}
\epsscale{1.0}
\plottwo{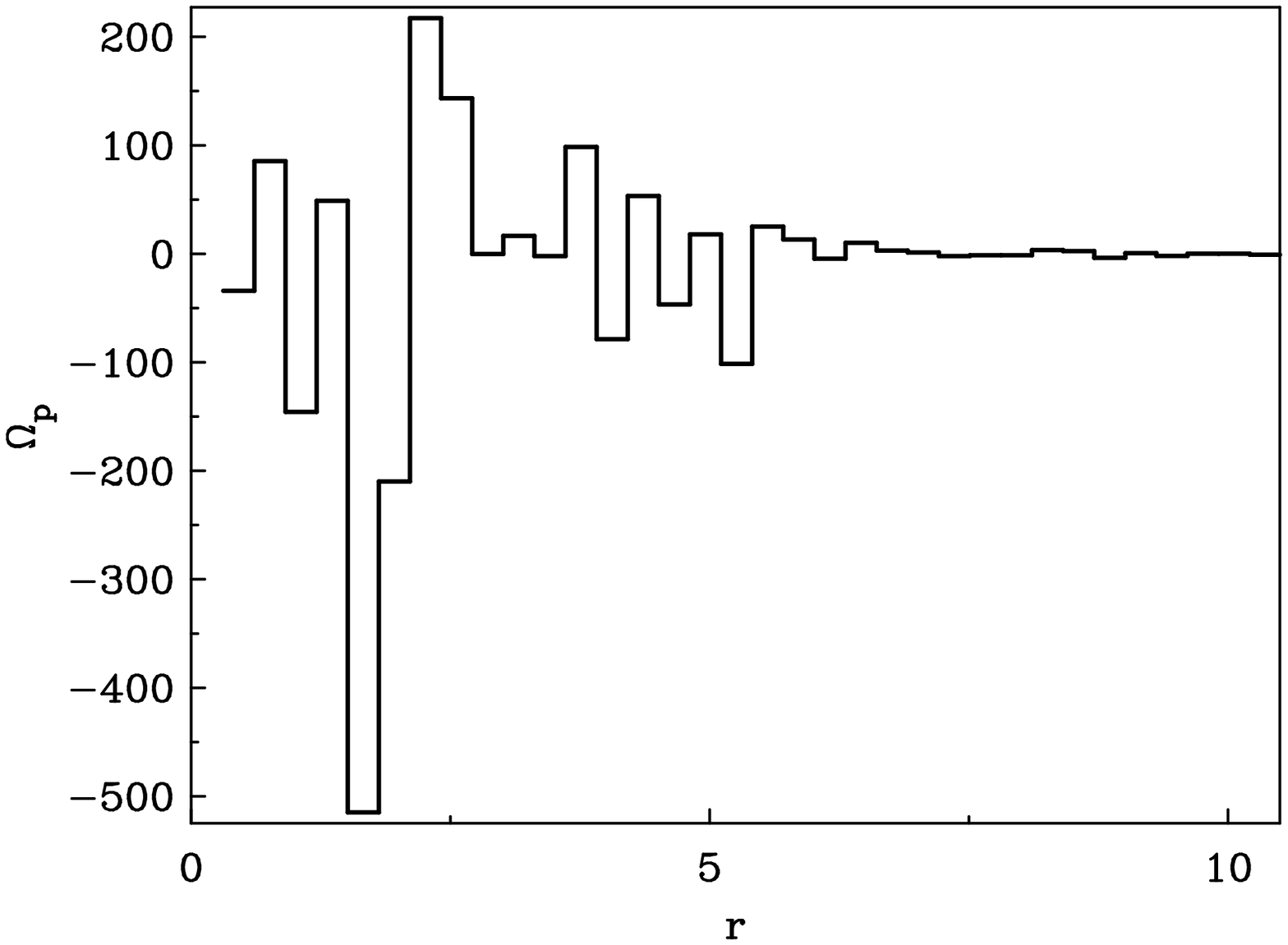}{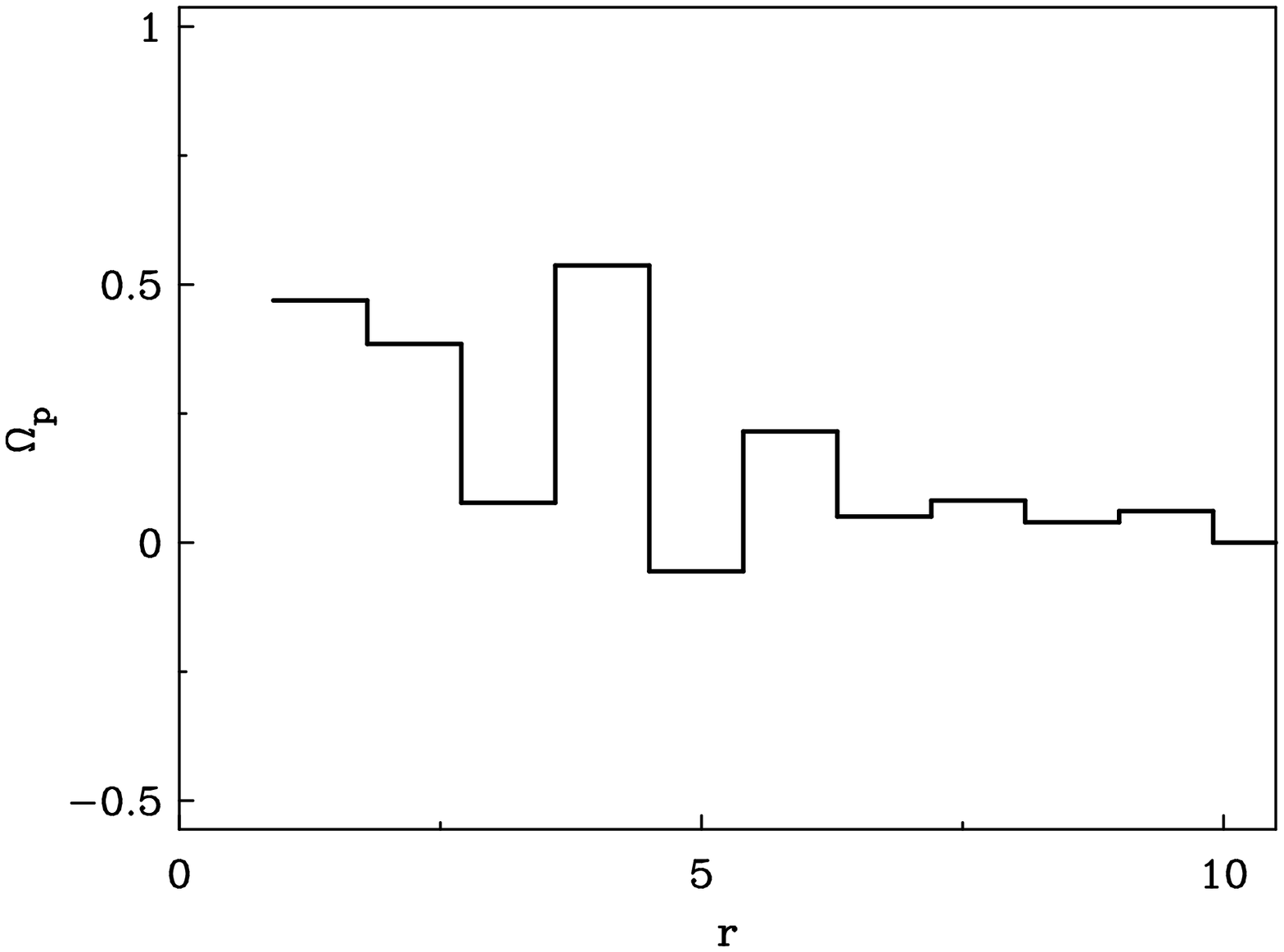} 
\end{center}
\caption{Plots of (unregularized) TWR solutions for two different binnings of data from the barred spiral simulation in $\S$ \ref{sec:SIM1} with SA=-45$\degree$ (see $\S$ \ref{sec:Nbody} for orientation convention).  The left (right) panel shows the solution generated using $\Delta r$=0.3 (0.9) bins.  }
\label{fig-unregularized}
\end{figure*}
Figure \ref{fig-unregularized} shows an example of (unregularized) TWR solutions for two different binnings using data from the barred spiral simulation of $\S$ \ref{sec:SIM1}.  By increasing the bin width the largest oscillations are reduced, but (as will become clearer) even tripling the bin width will not necessarily limit the propagation of noise to the level required for the extraction of realistic solutions.  (A more thorough discussion of the implications for TWR solutions will be deferred until the beginning of $\S$ \ref{sec:SIM1}.) 

Though the initial application of the TWR method on a real 
galaxy, namely NGC 1068 (MRM), showed little of the oscillatory behavior common to noisy, discretized 
Volterra-type solutions (outside $r$$\sim$1.5 kpc, anyway), the solutions 
were generated 
over relatively few bins (only five at most, over the region $r$=1.5 to 2.8 kpc).  In general, while large bin widths can often minimize the propagation of noise in the calculation, they can be expected to compromise solutions, as will be discussed in the sections to follow; naturally, the smaller the quadrature element, the more accurately the true radial variation of the pattern speed can be ascertained.  This is particularly critical as applied to disks sustaining multiple patterns where, as described later, a certain degree of radial precision is required for accurate separation of the pattern speeds.

Barring a large, limiting resolution, 
one should expect to be able to perform a sufficiently 
smooth quadrature wherein the number of elements in $\textbf{K}$ becomes large.  Since more elements in $\textbf{K}$ (and more bins over which 
to generate solutions) result in intrinsically noisy behavior--an effect most pronounced in the inner-most bins--gaining 
a finer, more accurate quadrature often means forfeiting control of the solution.  By combining regularization with the TWR calculation to force a smooth solution, 
however, one can counter this effect while maintaining the required precision.

Regularization also serves to alleviate the impact of non-global features that are most likely not included in the overall pattern (and which can singularly introduce large 
errors into the integrals).  Since rapid fluctuations in $\Omega_p(r)$ are penalized, discrepant points need not be 
avoided or ignored (as demanded in performing the TWR calculation on NGC 1068 in MRM).  While one may also use the 
alternative, which would be to fit models of $\Omega_p$ directly to equation (\ref{eq:twr}) and 
perform a grid-search to find the best model form and coefficients, we pursue regularization here, its speed making it 
preferred.  
\subsection{\label{sec:Regularization}Regularization}
Our procedure entails the following.  As a modification to the $\chi^2$ estimator minimized by solutions $\Omega_j$ of equation (\ref{eq:twr}), namely
\begin{equation}
\frac{\vert K_{ij}\Omega_j-b_i\vert^2}{\sigma_i^2}
\label{eq:chisq}
\end{equation}
with implicit sum over $i$ (and $j$) and errors $\sigma_i$ representing the measurement error of the $i^{th}$ data point 
$b_i$,
we introduce a regularizing operator, or smoothing functional $\textbf{S}$, containing a priori information in the manner of 
Tikhonov-Miller regularization (\citealt{tich}; \citealt{mill}) in which (in matrix form) solutions $\bf{\Omega}$ minimize
\begin{equation}
\vert \mathbf{\bar{K}}\cdot \mathbf{\Omega}-\mathbf{\bar{b}}\vert^2+\lambda \mathbf{\Omega}\cdot \mathbf{S}\cdot\mathbf{\Omega}.
\label{eq:tichonov}
\end{equation}
Here, the elements of $\bar{\textbf{K}}$ and $\bar{\textbf{b}}$ are $K_{ij}/\sigma_i$ and $b_i/\sigma_i$, respectively, and the role of 
$\lambda$--controlling 
the relative amount of $\chi^2$ minimization on the left to entropy maximization on the 
right--is explicit.

Reduced to a linear set of normal equations, this minimization returns smoothed solutions according to a 
modified version of equation (\ref{eq:twr}):
\begin{equation}
(\mathbf{\bar{K}}^T\cdot\mathbf{\bar{K}} +\lambda \mathbf{S}) \cdot\mathbf{\Omega} =\mathbf{\bar{K}}^T\cdot\mathbf{\bar{b}}
\label{eq:twrreg}
\end{equation}
Note that the regularizing functional, not necessarily upper triangular, introduces an anticipatory quality to 
solutions $\Omega_j$ whereby all bins at the same radius are coupled.  Furthermore, solving for components $\Omega_j$ no longer only involves a procedure like back-substitution, but requires rather an L-U decomposition (for instance) as well. 
\subsection{\label{sec:Smooth}The Smoothing Operator}
The real power in applying regularization to TWR calculations is in the freedom to choose how the smoothness 
of solutions is achieved.  For the purposes of distinguishing between different possible radial dependences for 
$\Omega(r)$, we choose $\textbf{S}$ to reflect a priori assumptions based on simple expectations from theory and observation.  Model solutions, then, each incorporating its own $\textbf{S}$, represent smoothed, testable realizations of the pattern speed.  These we restrict to simple forms in order to minimize the additional amount of information to be extracted from the data relative to the traditional TW method. 

For polynomial solutions, we consider only constant, linear and quadratic radial dependence. The 
elements of the 
smoothing $\textbf{S}$ are associated with the minimization of the $n^{th}$ derivative of $\Omega(r)$ for each 
polynomial solution of order $n$.
For instance, for linear solutions this entails minimizing 
\begin{equation}
\mathbf{\Omega}\cdot \mathbf{S}\cdot\mathbf{\Omega}  =\sum_{n=1}^{N-2}\vert -\Omega_{n}+2\Omega_{n+1}-\Omega_{n+2}\vert^2 
\label{eq:Slin}
\end{equation}
whereupon 
\begin{equation}
\mathbf{S}=
\left( 
\begin{array}{rrrrrrrrrrr} 
1&-2& 1& 0& 0& 0& 0& \ldots& 0 \\
-2& 5&-4& 1& 0& 0& 0& \ldots& 0 \\
1&-4& 6&-4& 1& 0& 0& \ldots& 0 \\
\vdots & \ & \ & \ & \ddots& \ &\ & \ & \vdots \\
0& \ldots& 0& 1&-4& 6&-4& 1& 0 \\
0& \ldots& 0& 0& 1&-4& 6&-4& 1 \\
0& \ldots& 0& 0& 0& 1&-4& 5&-2 \\
0& \ldots& 0& 0& 0& 0& 1&-2&1 
\end{array} \right).
\end{equation}

One may also choose a form for $\textbf{S}$ that identifies two or more distinct regions of independent radial behavior 
by invoking a step-function model.  
For the case of a barred spiral with a constant bar and quadratic spiral, for instance, this corresponds to minimizing
\begin{equation}
\mathbf{\Omega}\cdot \mathbf{S}\cdot\mathbf{\Omega}=\left\{
\begin{array}{ll}
\displaystyle\sum_{n=1}^{t-1}\vert \Omega_{n}+\Omega_{n+1}\vert^2 
&\mbox{for $n$$<$$t$} \\\\
\displaystyle\sum_{n=t}^{N-3}\vert -\Omega_{n}+3\Omega_{n+1}-3\Omega_{n+2}+\Omega_{n+3}\vert^2 
&\mbox{for $n$$\geq$$t$}.
\end{array}\right.
\label{eq:Sdiv}
\end{equation}
The elements of $\textbf{S}$ with $n<t$ reflect the a priori assumption that the bar pattern speed is constant, while those for 
$n>t$ associate a quadratically varying pattern speed with the spiral.  The index $t$, a free parameter, locates the radial bin where the 
transition between the two patterns occurs.  Obviously, the number of available bins constrains the order of the polynomial in a given radial zone.

Once we have chosen $\textbf{S}$, we initially choose $\lambda$ to reflect comparable amounts of $\chi^2$ minimization and 
regularization by letting $\lambda=\lambda_0=Tr\left\{\bar{\textbf{K}}^T \cdot\bar{\textbf{K}}\right\}/Tr\left\{\textbf{S}\right\}$.  Since we are in 
the business of generating solutions based on particular models, $\lambda$ is modified to arrive at the regularization 
required to return solutions of a given type.  This modification generally consists of an increase in $\lambda$ over $\lambda_0$.  
Consider how the regularizing parameter $\lambda$ regulates the degree of smoothness of the solution to the weight 
placed on the data: with $\lambda$=0, equation (\ref{eq:tichonov}) corresponds to $\chi^2$ minimization (and becomes an 
unbiased estimator with the smallest variance), however yielding highly oscillatory solutions, while $\lambda$$\rightarrow 
$$\infty$ corresponds to a maximally smooth estimator with non-vanishing variance.

Fitting data sets with different spatial coverage will change the effect of $\lambda$ on the solution (e.g. 
larger bins require less regularization).  The most appropriate choice for $\lambda$ (and $\textbf{S}$) should be made on a 
galaxy-by-galaxy basis, according to the quality of information to be extracted from observations. 

\section{\label{sec:Other}Other Considerations}
\subsection{\label{sec:Bias}Regularization-induced Bias}
By imposing assumptions about the smoothness of the pattern speed, regularization inevitably introduces complications 
for extracting realistic solutions. To understand how these arise, consider the solution for a barred spiral galaxy.
The nature of the calculation (from out to in) has implications for the accuracy of 
the bar estimates, in particular.  Not only do the $\Omega_j$ for bins covering the bar rely on the greatest number 
of matrix 
elements (and errors therein), but the bar estimate depends critically on the solution for the spiral and all outer bins via equation 
(\ref{eq:outin}).  Consequently, merely requiring the pattern speed in the outer bins to be constant with regularization out to the 
edge of the surface density (for instance)--effectively 
removing fluctuations that might better fit the data--will have consequences for the bar 
solutions.  So while the regularization is particularly fast and effective for tests for the radial behavior of 
patterns, it can also hinder the realization of accurate solutions.

For the simulations studied here, the risk of regularization-induced bias is inherited from the adopted quadrature.  Recall our requirement that all slices cover the 
full extent of the `emission'--so as to insure all integrals be fully converged--and relatedly, that the last matrix 
element governs the outermost slice position. In the barred spiral simulation of $\S$ \ref{sec:SIM1}, for example, such an extensive quadrature presents us with outermost slices that pass through a region where 
there is simply no discernible pattern (as indicated by the surface brightness distribution and its Fourier 
decomposition; see next section).  While these slices themselves do not provide direct estimates of the patterns of 
interest, the corresponding bin values are necessary for calculating the bar and spiral solutions.  Moreover, the quality of these solutions will be intimately related to the treatment of the outermost bins.  We therefore find that identifying, and reducing the influence of, the compromised zone by not enforcing regularization on these bins to be essential for accurate pattern speed measurement.
\subsection{\label{sec:Fourier}Fourier Diagnostics}
With the above concerns in mind, we have tested and used the following scheme.  Given slices that pass through an outer region which either contains little information from a strong pattern, is suspected of sustaining multiple patterns, or displays only faint emission, we choose in such a case to let the values in the outermost bins be calculated 
without regularization with the restriction, only, that they minimize the $\chi^2$.  Once a particular bin--at $r_c$, the cut radius--has been reached, regularization is imposed with all remaining inward bins generated accordingly.

In our procedure this corresponds to using an $\textbf{S}$ indexed by the cut bin $c$ that is lowest-block 
zero.  And the `cut' radius identifies the location in the disk where the outermost discernible pattern ends.  Note that this is in contrast to altogether ignoring the outer portion of the disk.  We prefer this procedure for two reasons: 1) as qualified later in $\S$ \ref{sec:TWTWR}, when each integral is truncated within the disk, the quadrature is at greater risk of ignoring information critical for characterizing the patterns uniformly throughout the disk and 2) like the transition radius $r_t$, we can easily incorporate $r_c$ as a free, though restricted, parameter in our models.

In practice, extracting the bar and spiral pattern speeds for the simulated barred spiral in $\S$ \ref{sec:SIM1} involves generating 
a group of solutions with 
various bar-to-spiral pattern transition radii for a given `cut' location.  Throughout the analysis, we choose the cut 
radius to reflect a priori knowledge of the outermost measurable pattern's termination radius estimated from the surface 
brightness and its Fourier decomposition.  When referred to, the power in each Fourier component, or mode $m$, is given by the norm of the complex Fourier amplitude
\begin{equation}
I_i=\sum_{n=1}^{N}e^{im\theta_n}
\end{equation}
where $\theta_n$ is the angular coordinate of each of the $N$ particles at each measured radius.

For the barred 
spiral in $\S$ \ref{sec:SIM1}, for example, we combine evidence from the surface density--where beyond the bar there 
is enhanced spiral surface density from only a limited radial zone--with the Fourier spectrum to identify a region in the 
disk outside the spiral which is susceptible to regularization-induced bias.  
Specifically, the Fourier power 
spectrum (Figure 
\ref{fig-BSpowerSEM}) shows that at $r$$\sim$5.0 the second clean hump in $m$=2 power decreases to almost zero, 
marking the end of the spiral.  Past this radius, the (strong) $m$=2 component (between $r$$\sim$6.0-8.0) is not associated with 
visible spiral structure (see Figure \ref{fig-SIMI}).  Reckoning this outer zone to be incompatible with a simple pattern speed model, then, we consider only the inward bar and primary spiral pattern speeds to be measurable with regularization.  Figure 19 of \citet{deb06} for this simulation 
(Figure \ref{fig-D06} in this paper) confirms this; not only does the spiral pattern terminate at $r_c$$\sim$5.0, 
but beyond this radius the pattern speed is multi-valued (this, of course, would be indiscernible in a real galaxy).  In this case, imposing form with regularization on bins of suspect 
quality and behavior 
outside the spiral will likely impair the solution of interest.  We therefore restrict the cut bin for the barred 
spiral simulation to 4.5$<$$r_c$$<$6.0, representative of where the primary spiral pattern terminates in 
the disk.  As mentioned in $\S$ \ref{sec:SIM1}, this step is substantiated by our finding that a cut radius of $r_c$=4.8 is one of several $\chi^2$ minima given a range of possible cut 
radii.  And furthermore, solutions generated in this manner are judged to overall provide a considerably better fit 
to the data than solutions where regularization is imposed out to the edge of the surface brightness (according to the 
scheme described in the next section).  
\subsection{\label{sec:weight}Weighting Schemes and Goodness-of-Fit}
Given the data, we simultaneously generate model solutions with different radial dependences for direct comparison 
for each side (y$>$0 or y$<$0) independently.  We then average the two like-model solutions together to construct a global solution. (In those instances 
when model solutions include a cut bin, the averaging occurs over the regularized part of the solutions only, in order 
to maintain the `unbiased' quality of the unregularized part of the solutions for each side of the galaxy.) 
Note that while the assumption that the patterns are indeed global is not overly inspired for the simulated galaxies studied here, 
putting this into practice on a real galaxy requires that the assumed galaxy kinematic parameters are accurate and that such symmetry exists.

With each global model solution we generate a complete set of $<$$v$$>$ using equation (\ref{eq:twr}).  We then judge each model through the $\chi^2_\nu$ ($\chi^2$ per degree of freedom) goodness-of-fit 
estimator of the reproduced to actual $<$$v$$>_i=b_i/(\int \Sigma dx)_i$ given measurement errors 
$\sigma_i^{<v>}$, in keeping with the standard TW analysis.  Note that with this choice the $\sigma_i$ in 
equation (\ref{eq:chisq}) are related to the errors $\sigma_i^{<v>}$ by $(\int \Sigma dx)_i$.  That is, the 
calculation fits to the $b_i$ given errors $\sigma_i$ while our $\chi^2$ estimator considers the differences from 
$<$$v$$>_i$ given errors $\sigma^{<v>}_i$. 

While 
for a real galaxy inaccuracies in the assumed position angle (PA) have the largest potential for introducing errors 
into $<$$v$$>_i$, we prefer that the measurement errors  $\sigma^{<v>}$ reflect random noise in the data, only.  (Systematic 
errors prove more practically assessed through direct tests of the sensitivity of the results to departures from the nominal 
values for the PA or inclination, for instance.)  For the simulations studied here, then, we obtain errors $\sigma_i^{<v>}$ 
under the assumption that the inverse mirror image of each $<$$v$$>_i$ on one
side of the galaxy should be the same as on the other side (i.e. the patterns are symmetric).  We then assign a global error $\sigma^{<v>}$ to 
each slice where $\sigma^{<v>}$=$\left(\sum_{i=1}^{2N}\sigma_i^{<v>}\right)$/2N (and $N$ is the 
number of bins/slices used in the TWR calculation on a single side).

In practice, the simplest $\chi^{2}$ weighting schemes are either uniform weighting for all slices or 
weighting by the intensity, which should give more weight to slices where the signal is strongest.  We have chosen the 
former since we are interested in $\Omega_p$ over a broad radial range, and prefer that our result is not dominated by 
just the slices with the highest signals (which can vary dramatically).  Furthermore, the choice of assigning an identical error $\sigma^{<v>}$ 
to each slice carries with it an implicit weighting scheme for equation (\ref{eq:tichonov}).  For an exponential surface 
brightness profile, for example, slices on either side of the galaxy will have progressively smaller $\int \Sigma dx$ 
as $\vert y_i \vert$ increases.  This corresponds to errors $\sigma_i$ proportional to $\int \Sigma dx$, then, that grow 
larger from out to in.  Since in most cases the uniform weighting scheme will be in actuality most restrictive of the 
outer bins--the  goodness of which will affect the solution inward--this choice is particularly well suited for the TWR 
calculation.

Given our choice of weighting scheme, there are two important considerations which demand that we calculate the $\chi^2$ over all slices. First, the 
inner-most bins contribute to the $<$$v$$>$ in only the inner most slices and hence contribute relatively negligibly to the 
$\chi^2$, despite possibly larger weights $K_{ij}$ (reflective in part of a surface brightness that is centrally peaked, say) 
than bins at larger radius.  This is especially true when an inner pattern appears only over a small fraction of the 
total bins, and the reproduced $<$$v$$>$ of even those slices that pass directly through the inner pattern still rely (perhaps 
predominantly) on the solution out to the largest radial bin. However, since the goodness of the inner bins is directly 
related to the goodness of the bins at larger radii, by considering all slices we more effectively judge the whole 
solution.

Secondly, in cases when some number of outermost bins are calculated 
without regularization, it is critical to account for the (largely positive) effect that these bins have on the solution inward, 
especially from model to model.  In our current scheme, the values of the `unregularized' bins are not quite identical to 
those in the completely unregularized 
solution since the quantity that they minimize still includes participation from non-zero elements in the smoothing 
functional $\textbf{S}$ and what is currently a model-dependent $\lambda$.  That is, slight variation in the values of the `unregularized' bins from model to model is apparent, and we 
cannot ignore the minor differences this introduces to the regularized part of the solution.  Though this is a minor effect, 
by considering the $<$$v$$>$ of all slices in the $\chi^2$ we prevent solution-preference based on the `unregularized' bins 
(which act essentially as stand-ins) from being introduced.

The first consideration above also prevents us from calculating judicious error bars on the solutions according 
to the variation of individual model parameters over a typical $\chi^2$ confidence interval; in practice, the 
value of an inner pattern with minimal radial extent can change considerably with little effect on the $\chi^2$.  
Indeed, we find that the errors generated according to such a prescription are unrepresentative of the goodness of the 
solutions as returned by the calculation. 

In $\S$ \ref{sec:iError}, we describe the dependence of both 
inner and outer speeds on the location assigned to the transition between the two.  An obvious progression for future 
applications of the TWR method would be an exploration of the covariance of what we consider here, to first order, 
`free parameters', especially for the purposes of improved error estimation.  Presently, however, we construct error 
bars for model solutions by considering 
the range of parameters in the best solutions at different assumed projections.  As will pertain to the sections 
that follow, by considering an overall solution in this manner we can fairly account for the uncertainty introduced 
for real galaxies by the reality that each can be sampled at only a single PA.   
\section{\label{sec:SIMS}Tests of the TWR Method on Simulations} 
In order to establish guidelines for applying the regularized calculation to observations of real galaxies, we next perform tests of the method on simulations with known pattern speeds.  Each case invokes unique models for $\Omega_p(r)$ which we motivate and discuss in detail.  The procedure for engaging the method with maximum accuracy then follows from careful examination of the quality of solutions given the available information.  Though quite detailed, these individual studies together constrain general scenarios and practices to extrapolate onto observations of similarly structured, real galaxies.  
\subsection{\label{sec:Nbody}N-body Systems}
We use three simulations in this study.  The first,
which we refer to here as simulation I, constitutes a barred galaxy
with spiral structure.  Originally presented in \citet{deb06} (hereafter D06) where it is referred to as run L2.t12, it 
consists of a live disk immersed in a rigid halo.  A
complete description of the model parameters are contained in D06.

The second (spiral) simulation, which we refer to here as simulation II, is
unpublished.  It was designed with the main aim of generating strong
spiral structure using the groove mode mechanism of \citet{sl89} and \citet{sk91} in which dynamical instability develops from a 'groove', or narrow feature, in the phase-space density at a particular angular momentum.  A trailing spiral wave is generated, and at the Lindblad Resonances of the wave, further grooves develop such that the instability is recurrent.  Like simulation I, it consists of a rigid halo and live disk, but also includes a live bulge component.  The bulge constitutes $25\%$ of the
baryonic mass and is sufficiently concentrated that a bar is very slow
in forming.  The disk has Toomre-$Q = 1.2$; in order that a strong
spiral was seeded, $\sim 6\%$ of disk particles in a narrow angular
momentum range were removed leaving 4$\times$10$^6$-169480 particles including the bulge.  The result, as can be seen in Figure \ref{fig-SIM2}, is the formation of a strong but transient spiral.

The last simulation in
this paper, simulation III, is a double-barred galaxy generated using
the method of \citet{dshen07}.  This high-resolution simulation consists of live
disk and bulge components in a rigid halo potential. The model has
$\sim 4.8$ million equal mass particles, with $\sim 4$ million in the
disk and $\sim0.8$ million in the bulge such that the bulge has mass
$M_b=0.2M_d$, where $M_d$ is the disk mass. The initial Toomre-$Q$ of
the disk is $\simeq2$. The formation of the secondary bar is induced
by making the bulge rotate (to mimic a pseudobulge). More details of
the simulation can be found in \citet{sdeb07} where it is referred to as run D.

As in D06, all lengths and velocities are here presented in natural units.  We analyze a snapshot of each simulation at 
a single time step with the disk in the $xy$ plane.  By rotating the system about the $z$ axis we assign a 
line-of-sight direction to establish the kinematical major axis.  Another rotation about the $x$ axis gives the system 
an inclination $\alpha$ (chosen throughout at $\alpha$=45$\degree$, unless otherwise specified).  The snapshot is then 
projected onto the sky-plane where $x_{obs}$=$x$ and $y_{obs}$=$y\cos{\alpha}$.  For a given slice spacing, the slices along which the calculations occur are aligned perpendicular 
to the line-of-sight direction (parallel to the kinematical major axis).  The orientation of these slices, which is identical to the disk PA in a real observation, is designated uniquely the slice angle (SA) in the studies that follow.

\subsection{\label{sec:SIM1}Simulation I: Barred Spiral Galaxy}
\begin{figure}
\begin{center}
\epsscale{.90}
\plotone{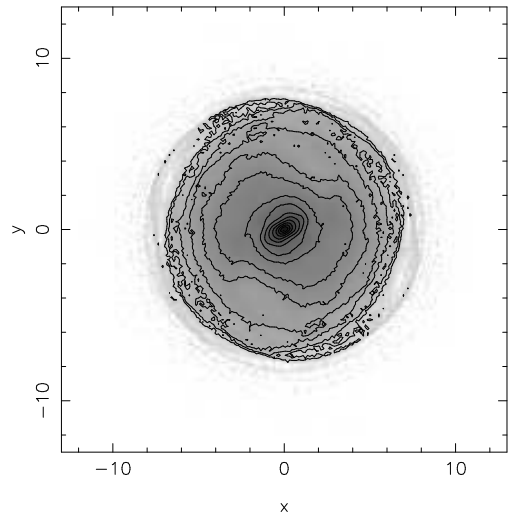}
 \end{center}
    \caption{Face-on display of the barred spiral simulation's surface brightness distribution projected with a -30$\degree$ rotation about the $z$ axis. For reference, the alignment of the TWR quadrature for a frame at this orientation is designated SA=+60$\degree$.
\label{fig-SIMI}}
\end{figure}
 \begin{figure}
\begin{center}
\epsscale{.90}
  \plotone{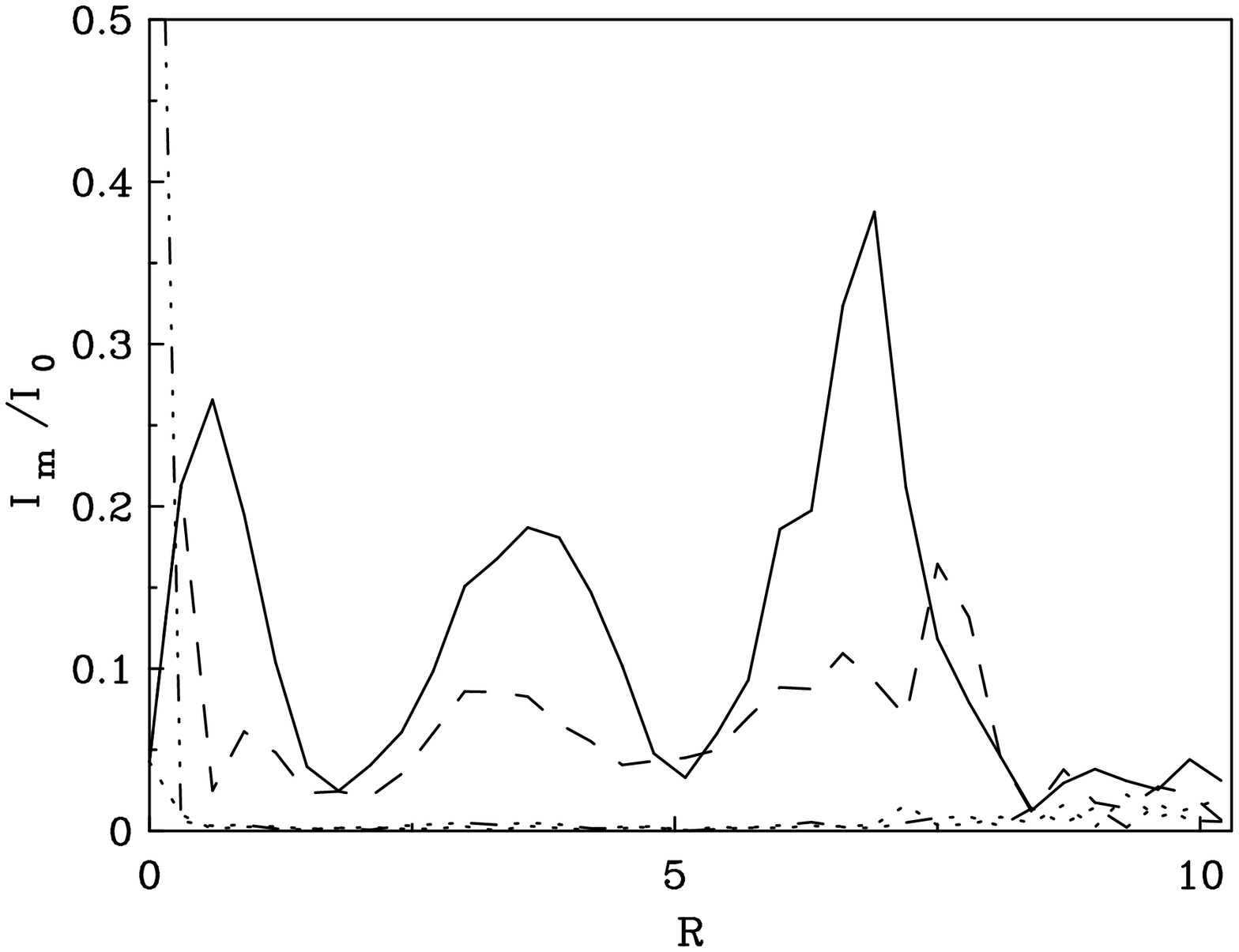}
 \end{center}
    \caption{Fourier power spectrum of the barred spiral simulation's surface brightness distribution shown in Figure \ref{fig-SIMI}.  Modes up 
to $m=4$ are plotted as a function of radius with lines for $m=1$ in dot, $m=2$ in solid, $m=3$ in dash-dot-dot, and $m=4$ in dash.\label{fig-BSpowerSEM}}
\end{figure}
\begin{figure}
\begin{center}
\epsscale{.90}
 \plotone{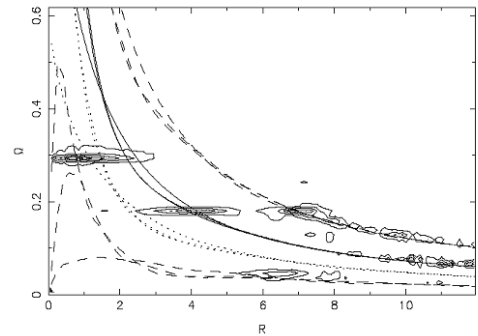}
 \end{center}
    \caption{Contours of the barred spiral simulation's $m$=2 Fourier mode showing a bar pattern
    speed $\Omega_b$=0.29, a bar-to-spiral transition of $r_t$$\sim$2.5, and a dominant spiral pattern
    speed $\Omega_s$=0.18 out to $r$$\sim$4.5-5.5 beyond which multiple spiral modes exist.
\label{fig-D06}}
\end{figure}
The bar and spiral structure in this simulation, first presented in D06, is featured out to $r$$\sim$5.0, clear in the surface 
density (Figure \ref{fig-SIMI}) and its Fourier decomposition (Figure \ref{fig-BSpowerSEM}).  Beyond $r$$\sim$5.0, the Fourier decomposition indicates the possible presence of a third pattern.  With step-models for $\Omega_p(r)$, then, we might reasonably extract pattern speeds for three distinct structures.  However, the $m$=2 mode between 5.0$<$$r$$<$8.0 is not associated with a strong surface density enhancement.  And as remarked upon in $\S$ \ref{sec:Fourier}, Figure 19 in D06 reproduced here in Figure 
\ref{fig-D06} shows that the pattern in this radial zone is maintained by multiple distinct pattern speeds.  (Note that a real galaxy would not be disposed to the analysis provided with this type of plot.  It is available for this simulation, only, and we include it here for the sake of comparison).  In a clear account of regularization-induced bias, test solutions based on a three-pattern speed model have considerably larger $\chi^2$ than those parameterizing a bar and single spiral.  We therefore reject models with a third pattern speed and use our `cut' scheme where 
solutions are generated without regularization up to a cut radius $r_c$ which parameterizes the 
end of the primary spiral pattern.

The actual pattern speeds of the bar and spiral structure to be reproduced by our solutions are $\Omega_b$=0.29 for the bar, and a constant $\Omega_s$=0.18 for the spiral, as estimated from Figure \ref{fig-D06} (from which we also estimate a bar-to-spiral pattern transition radius $r$$\sim$2.5).  In our models of $\Omega_p(r)$, we express the a priori assumption that the bar pattern speed is constant by using an $\textbf{S}$ like that in 
equation (\ref{eq:Sdiv}) but where, for $n>t$, $\textbf{S}$ is reflective of either a constant, linear, or quadratically varying spiral pattern speed.  
Solutions with spirals of order 0, 1, and 2, then, have a total of 4, 5, and 6 degrees of freedom, respectively.  The free parameter $t$ we restrict for all models such that 1.8$<$$r_t$$<$3.0, according to bar-length estimates from Figure \ref{fig-SIMI} ($a_B$$\sim$2.2) and Figure \ref{fig-BSpowerSEM}.

For $\Delta r$=0.3 bins, we require a total of 71 slices (35 on each side) to reach the edge of the surface brightness at 
$r$$\sim$10.5.  This places the cut bin between 15$<$$c$$<$20, according to the previously motivated restriction 4.5$<$$r_c$$<$6.0 estimated from Figures \ref{fig-SIMI} and \ref{fig-BSpowerSEM}.

\begin{figure}
\epsscale{.90}
\begin{center}
\plotone{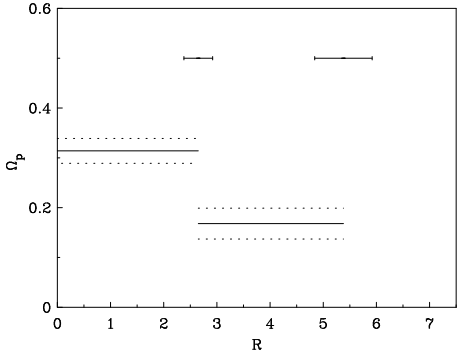}
\end{center}
    \caption{The best-fit solution and error bars from the TWR method applied to the barred spiral simulation averaged over six SAs.  The bar 
$\Omega_b$=0.31$\pm$0.02 and spiral $\Omega_s$=0.17$\pm$0.03 are shown as solid lines with dashed errors. Errors in the bar-to-spiral transition $r_t$=2.6$\pm$0.28 and spiral termination radius $r_c$=5.38$\pm$0.54 are represented by horizontal error bars at the top. 
\label{fig-BSsol}}
\end{figure}
In light of the discussion in $\S$ \ref{sec:weight}, we construct errors for our estimates to reflect the expected accuracy of TWR bar and spiral solutions given a particular observational 
scenario.  Specifically, we perform the calculation for a range of SAs spanning the upper half-plane of the galaxy (quadrants I and 
II), namely $\pm$15$\degree$, $\pm$45$\degree$, and $\pm$75$\degree$.  Each SA corresponds to a unique disk PA.  The resultant bar and spiral estimates 
generated using the $\Delta r$=0.3 bin width (to be discussed at length in the following sections) are listed in Table \ref{tab-BSall} and the average and rms of the best-fit
solutions for this SA range are shown in Figure \ref{fig-BSsol}.  There, horizontal error bars represent the dispersion in 
$r_t$ and $r_c$ 
in the solutions based on variations from SA to SA for the $\Delta r$=0.3 radial bin width.

\begin{figure*}
\begin{center}
\epsscale{1.0}
\plottwo{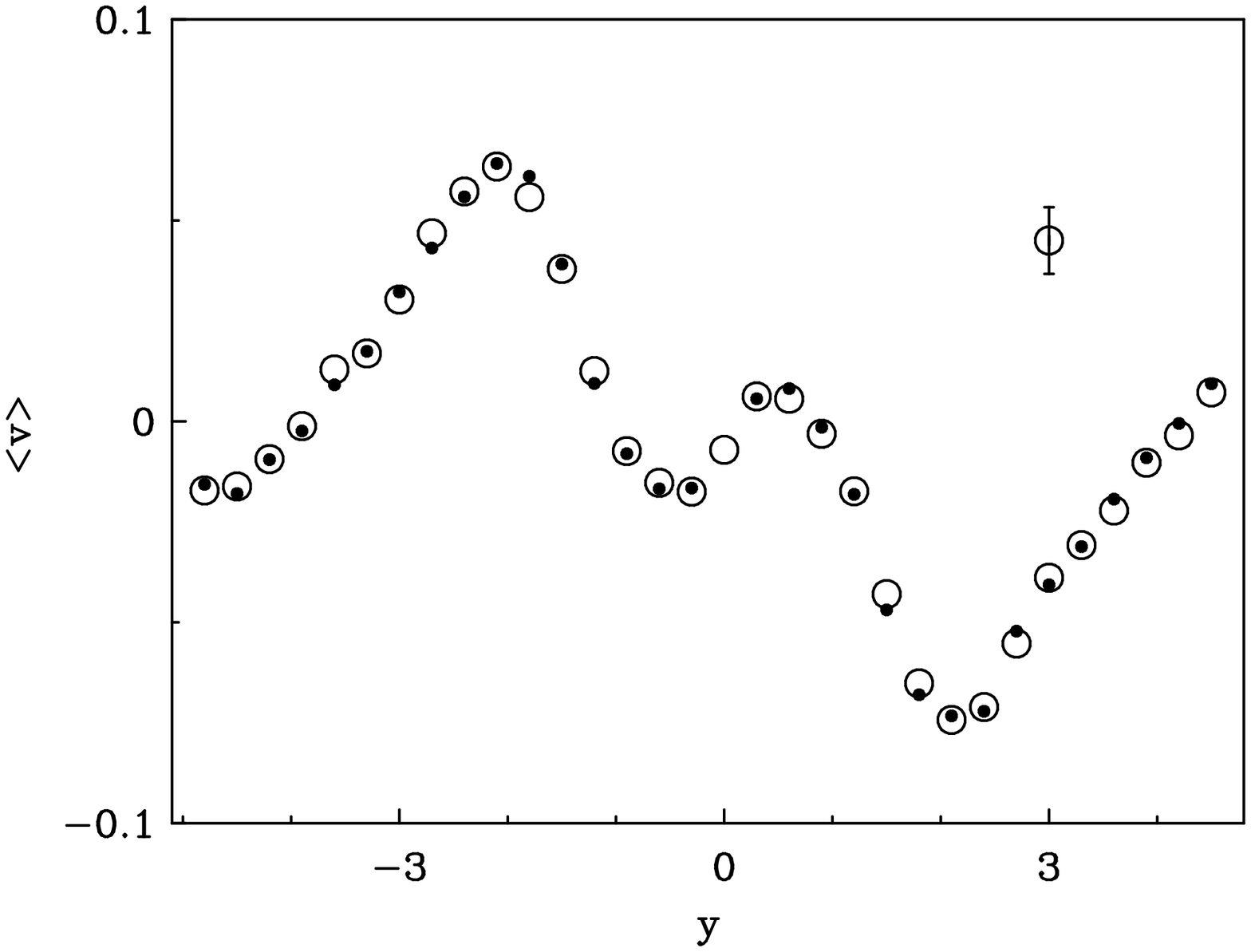}{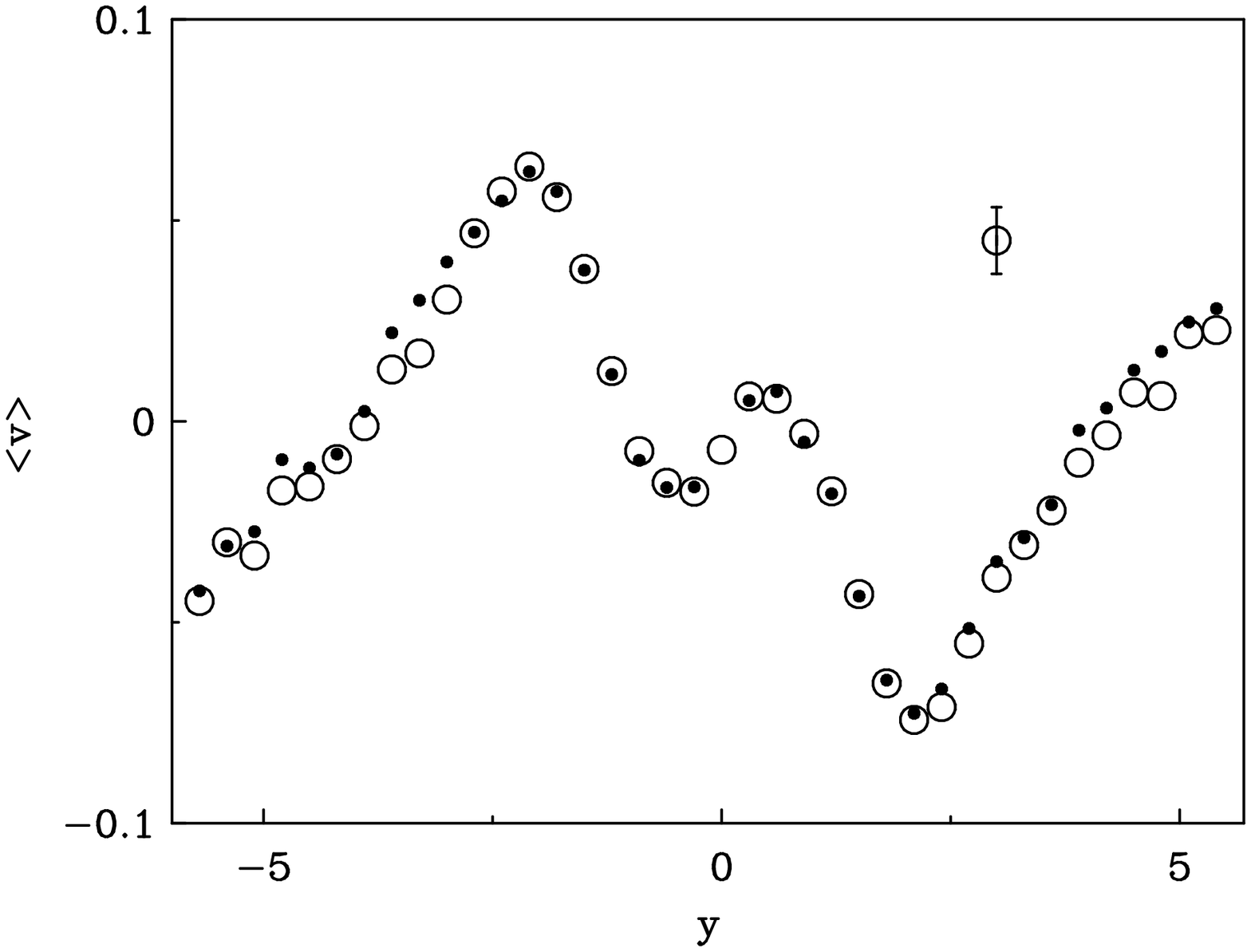} 
\end{center}
\caption{Comparison of model solution-reproduced (closed circles) to actual (open circles) integrals 
$<$$v$$>_i$=$b_i$/$\int \Sigma dx$ as a function of slice position $y$ for the barred spiral simulation for a). the best-fit constant $\Omega_b$, constant $\Omega_s$ 
solution, with $r_t$=2.4 and $r_c$=4.5 and b). constant $\Omega_b$, quadratic $\Omega_s$ solution with $r_t$=2.7 and $r_c$=5.4 for SA=15$\degree$.  Only those slices which show 
a contribution from bins inward of $r_c$ are shown. The adopted global error $\sigma^{<v>}$ is shown in the upper right.}
\label{fig-vslice}
\end{figure*}

Rewardingly, the best-fit solutions are quite accurate; the comprehensive spiral and bar estimates in Figure \ref{fig-BSsol} are 6.7\% and 8.3\% from their actual values (with $\sim$8\% error in $\Omega_b$ but a slightly larger error of $\sim$18\% in $\Omega_s$).  So, too, are the determinations of $r_t$ and $r_c$, according to Figures \ref{fig-BSpowerSEM} and \ref{fig-D06}.  Furthermore, our solutions correctly reproduce the functional form of the spiral pattern speed.  At all SAs, out of all solutions with $r_t$ and $r_c$ within their restricted ranges, the lowest reduced $\chi^2$ solution corresponds to a constant spiral.  Figure \ref{fig-vslice} 
shows a comparison between the actual and best-reproduced $<$$v$$>$ at each slice position $y_i$ used in the calculation 
for the 15$\degree$ slice angle.   The $<$$v$$>$ reproduced by the best-fit constant bar and constant spiral solutions are shown 
in the left panel, while those from the optimum (lowest-$\chi_\nu^2$) solution with a quadratic spiral are plotted in the right.

The close reproduction of the actual pattern speeds by the solutions in Figure \ref{fig-BSsol} occasions further evaluation of strictly unregularized TWR calculations.  Consider the values in radial bins inside $r$$\sim$2.0 in typical unregularized solutions for this simulation (Figure \ref{fig-unregularized}).  That there is little to no indication of $\Omega_b$=0.29 in the left plot is perhaps not surprising: the large number of bins which accompany the choice of the small bin width would seem to guarantee a high level of noise propagated throughout the solution.  However, in the slightly more stable solution with the wider bin width, the inner bins are still unrepresentative of the actual pattern speed in this zone.  We can understand this as a systematic error introduced by the noise which not only propagates but also compounds as the full solution assembles from the outermost to the innermost radius; $\Omega_p(r)$ in a given bin reflects errors from all exterior radial bins making the value in that bin more likely far removed from the actual value.
\begin{table}
\begin{center}
\caption{TWR estimates for Simulation I. \label{tab-BSall}}
\begin{tabular}{rcccc}
\\
\tableline\tableline
SA&\textbf{$\Omega_b$}&\textbf{$\Omega_s$}&$r_t$&$r_c$\\
\tableline
75$\degree$&0.325&0.149&3.0&5.7\\
45$\degree$&0.327&0.111&2.4&4.8\\
15$\degree$&0.343&0.184&2.4&4.5\\
-15$\degree$&0.269&0.172&2.7&4.8\\
-45$\degree$&0.313&0.199&3.0&5.4\\
-75$\degree$&0.303&0.203&2.4&6.0\\
\tableline
-&0.29&0.18&2.5&5.0\\
\tableline
\end{tabular}
\tablecomments{TWR bar and spiral pattern speeds from the barred spiral simulation listed here are estimated with TWR solutions calculated using a $\Delta r$=0.3 bin width for a range of SAs.  The third and fourth 
columns list the connate estimates for $r_t$ and $r_c$.  Values for the actual pattern speeds are shown in the last row.}
\end{center}
\end{table} 
\subsubsection{\label{sec:iError}Morphology-dependent Effects and Intrinsic Limitations}
The use of simulations which can be studied at multiple projections provides us with perhaps the most critical 
assessment for the accuracy of TWR solutions.  Figure \ref{fig-BSsol} suggests that the TWR method should perform well 
for any given viewing angle.  However, though still quite small, the rms in each 
estimate is largely reflective of the non-trivial effect that the orientation of 
the pattern with respect to the slice angle used in the calculation can have on solutions.

We can understand the origin of the differences in solutions for the range of SAs in 
Table \ref{tab-BSall} by considering the impact of the limited azimuthal 
range of the bright spiral enhancement (clear from the surface brightness distribution in Figure \ref{fig-SIMI} where 
the spiral 
extends almost perpendicular to the bar major axis).  That is, at all slice angles the quadrature accumulates fragmentary information from the spiral since only some of the slices that cross the full radial zone of the spiral pattern intersect the strong spiral structure.  But whereas both the bar and spiral estimates
seem to suffer at SAs in quadrant II (i.e. positive SAs), our measurements of $\Omega_s$ in quadrant I are quite 
accurate.  According to the morphology, in quadrant II it appears that the limited sampling of the spiral asymmetry implicit in slices other than those that also pass through the bar entails slightly less accurate spiral estimation.

To interpret the distinction between solutions from the two quadrants, consider the combined influence of 
regularization and our chosen weighting scheme.  Specifically, since the regularizing $\textbf{S}$ induces the 
coupling of all bins within the same radial zone, even 
when the spiral-zone crossing slices do not intersect the strong 
spiral enhancement, the bins there inherit information from bins at the same radius from slices which do intersect the 
arms.  According to our weighting scheme, however, this coupling is not uniform; the degree of support at each azimuth 
is influenced by the measurement errors for each slice which grow larger from out to in.  
As a result, $\Omega_s$ is 
best constrained by information from slices passing through the outer radial zone of the spiral alone (not those 
passing through the bar).  So when in quadrant II the spiral asymmetry appears in only the inner slices, $\Omega_s$ 
is less precise then when these outer slices clearly intersect the spiral arms.  According to equation (\ref{eq:outin}), 
since the bar estimate is directly related to that of the spiral, the result for these SAs is error in both $\Omega_s$ 
and $\Omega_b$.

The corresponding determination for the radial domain of the bar pattern, on the other hand, is not as 
obviously sensitive to this issue.  Not only is the bar-end reasonably well defined in both quadrants, but information from the bar which contributes to the parameterization of $r_t$ is reinforced with regularization in the manner described above.  The spiral termination radius, too, seems fairly consistent from SA to SA.  But since the asymmetry is weaker (and there is less information) at that location in the disk, we find that this parameter requires the most restriction (indeed, our determinations of $r_c$ completely span the allowed range).

Overall, then, our determinations for $r_t$ and $r_c$ are stable and accurate, each with less than 11\% error, even in quadrant II.  Nevertheless, since the pattern speed estimates from SAs in quadrant I seem to comparatively benefit from the high quality of information from both patterns, we conclude that position angles which provide the most uniform slice coverage of all patterns in the disk are preferred.

The largest disparity between step-model solutions from various slice angles can largely be attributed to limitations 
in determining the location of the transition between the two patterns.  That is, solutions are affected by the finite bin width inherent to the numerical calculation; slight incompatibility between the actual transition and that to which the solution is limited (given the 
bin width) can result in errors around 10\%.  This is a more pervasive effect than morphology alone and more obvious 
with the use of a slightly larger bin width than $\Delta r$=0.3.  Nonetheless, we can make several informed inferences about the result of the finite bin width by considering the nature of the TWR calculation.  Specifically, since the transition determines the contributions of inner and outer patterns to the integral on the right hand 
side of equations (\ref{eq:twint}) and (\ref{eq:twr}) through the matrix elements $K_{ij}$, a mismatch between the transition 
bin and the actual transition radius will corrupt the separation of the contributions of the two patterns. Note that this is 
precisely the source of the covariance between model parameters intimated in $\S$ \ref{sec:weight}.

To understand the effects of (minor) radial pattern misassignment, consider a 
step-function solution that parameterizes the extent of a bar pattern speed along with that of a spiral.  When the transition bin between the two patterns underestimates the actual transition by a fraction of the bin width, for instance, we would expect
the numerical calculation to effectively subtract off from $b_i$ a contribution from the (lower) outer pattern where a 
higher 
pattern actually exists (e.g. eq.[\ref{eq:outin}]), slightly raising the value of the inner pattern speed.  Conversely, we expect an overestimation of the transition to result in a slightly lower inner pattern speed.  More subtle effects can occur, however, depending on the geometry of the patterns, as illustrated for this simulation in $\S$ \ref{sec:Trans}.

For a range of SAs, we emphasize that the resultant errors--in possible combination with an undersampled 
transition--are minimal, as long as the bin size is sufficiently small, and are not the result of vastly different 
transitions for each SA;  we find that the 
transition between patterns in the best solutions is generally relatively stable with changes in slice angle.  
This may 
be unexpected from the perspective of the traditional TW method since the projected length presented to the slices by 
the bar will depend on the relative orientation of the slices with its major axis (and  
slice-orientation errors tend to be large, as detailed next for this simulation).  
\subsubsection{\label{sec:TWTWR}TWR vs. TW}
\begin{figure}
\begin{center}
\epsscale{.90}
\plotone{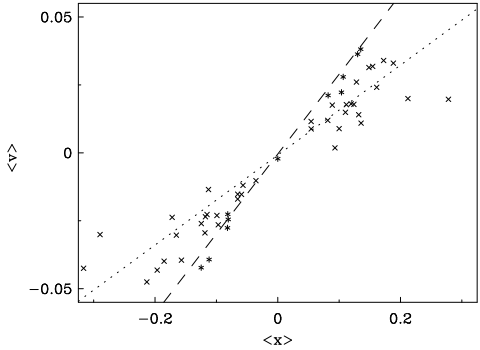}
 \end{center}
    \caption{Plot of $<$$v$$>$ vs. $<$$x$$>$ for all slices with $\vert y \vert$ $<$5.1 for the barred spiral simulation at SA=-45$\degree$.  The dashed line is the best-fit straight line to the inner eleven slices (stars), while the dotted line is the best-fit straight line
to all slices shown (stars and x's).  
\label{fig-tw45}}
\end{figure}

In light of the above results, we next examine the improvements available to pattern speed estimation using the TWR calculation relative to the TW method.  We specifically compare the bar pattern speed estimates arrived at using the 
TWR method with those using fully extended TW integrals.  Though TW estimates of $\Omega_b$ in the presence of a 
secondary structure may also be attempted using truncated integrals that extend to just past the end of the bar (such that information from the bar alone is dominant), 
we examine the former case for two reasons: 1) to compare the two methods under identical conditions (i.e. 
using the same data points $<$$v$$>_i$) and 2) to study the influence of the relatively weak spiral 
(and evaluate the assumption of negligible non-axisymmetric motions beyond the bar).
\begin{table}
\begin{center}
\caption{Optimal TW bar estimates for Simulation I. \label{tab-BStw2}}
\begin{tabular}{rccc}
\\
\tableline\tableline
SA&$N$ slices&\textbf{$\Omega_p$}&$\sigma$\\
\tableline
75$\degree$&7&0.270&$\pm$0.010\\
45$\degree$&11&0.282&$\pm$0.062\\
15$\degree$&17&0.208&$\pm$0.007\\
-15$\degree$&15&0.248&$\pm$0.026\\
-45$\degree$&11&0.294&$\pm$0.008\\
-75$\degree$&9&0.233&$\pm$0.053\\
\tableline
\end{tabular}
\tablecomments{All entries originate through the use of an optimal number of slices spaced at $\Delta y$=0.3.  The number of slices $N$ used in the TW calculation are indicated.}
\end{center}
\end{table} 
Though we use fully extended 
integrals to perform the TW calculation--even when making estimates of the 
bar pattern speed--the innermost slices clearly supply evidence for a bar pattern speed 
 that is distinct from that of the other structure in the disk.  Figure \ref{fig-tw45} shows a typical plot of $<$$v$$>$ vs. $<$$x$$>$ for this simulation where the inner eleven slices are indeed best fit by a steeper slope than for all slices. 

For all of the other SAs studied in the previous section, we measure $\Omega_p$ with a unique number of bar-crossing slices.  This is intended to reproduce an optimal TW observing strategy that makes use of only those slices that intersect the enhanced bar surface density.  That is, since the projected length presented to the slices by the strong bar structure depends on the relative orientation of slices with the major axis of the bar, the number of slices that intersect the bar enhancement varies from SA to SA (from seven to 17 for this simulation at the six studied SAs).  Table \ref{tab-BStw2} lists this optimal number of slices $N$ at each SA along with the corresponding pattern speed estimate.  All entries in the table correspond to slopes of best-fit straight lines (and the corresponding intrinsic scatter) in plots of $<$$v$$>$ vs. $<$$x$$>$.
\begin{table}
\begin{center}
\caption{Traditional TW estimates for Simulation I. \label{tab-BStw}}
\begin{tabular}{rcccc}
\\
\tableline\tableline
\multicolumn{1}{c}{SA} &\multicolumn{2}{c}{inner 17 slices}&\multicolumn{2}{c}{all slices}\\
\cline{2-5}
 &\textbf{$\Omega_p$}&$\sigma$&$\Omega_p$&$\sigma$\\
\tableline
75$\degree$&0.234&$\pm$0.004&0.115&$\pm$0.008\\
45$\degree$&0.217&$\pm$0.045&0.094&$\pm$0.018\\
15$\degree$&0.208&$\pm$0.007&0.157&$\pm$0.009\\
-15$\degree$&0.216&$\pm$0.016&0.121&$\pm$0.007\\
-45$\degree$&0.278&$\pm$0.009&0.166&$\pm$0.009\\
-75$\degree$&0.175&$\pm$0.010&0.128&$\pm$0.010\\
\tableline
\end{tabular}
\tablecomments{As in Table \ref{tab-BStw2}, the TW estimates for this SA range are generated with slices spaced at $\Delta y$=0.3.  Here, the second column lists the estimates $\Omega_p$
along with errors $\sigma$ using the inner seventeen slices, respectively, while the last column is from a fit to 
all slices with $\vert y \vert$$\leq$5.0, out to the inferred spiral-end.}
\end{center}
\end{table} 
Upon inspection, Table \ref{tab-BStw2} seems to suggest that even in the presence of the spiral asymmetry information from the bar is maximal in the bar-crossing TW integrals.  Despite also reflecting a contribution from the spiral 
pattern, these bar estimates are fairly accurate (though presumably not as accurate as would be the case 
for a strictly SB0 galaxy such as NGC 7079). However, as in Table \ref{tab-BStw}, if we extend the slice coverage at each SA out to $\vert y\vert$$\sim$2.4 using the inner 17 slices (at $\Delta y$=0.3 spacing)--closer to the full extent of the radial zone of the bar pattern, according to Figure \ref{fig-BSsol}--then the quality of the TW estimates diminishes with the inner, bar 
pattern speed estimates approaching that from a fit to all slices with $\vert y \vert$ $<$5.1; even the true bar-crossing slices contain non-negligible information from the spiral pattern.

Unlike the TW method, the TWR method is not relegated to the use of only those slices where the bar contribution is maximized.  In principle, the inner solutions are accessible through
the very use of information from beyond the bar (namely from the zone of the spiral), and are improved when this
information is radially coupled (e.g. by regularization).  This aspect of the calculation allows for the return of pattern speed solutions 
without reference to an assumed pattern extent, and provides, moreover, an independent means of determining the radial 
domain of patterns.

Of course, this is not to imply 
that a deficiency of usable slices in TW estimates prevents accurate pattern speed measurement, or determination of pattern extent, 
for that matter; using slices that cross primarily through the enhanced emission from the bar, TW estimates from 
SB0 galaxies have been successfully used to observationally 
confirm that bars end at or inside their corotation radii.  However, it does suggest that in the presence of non-negligible asymmetry exterior to the bar, TW bar estimates are susceptible to errors introduced by the use of slices positioned near the bar end which are presumed to reflect the bar pattern speed but in reality include a significant contribution from this structure.  Consider a typical plot of $<$$v$$>$ vs. $<$$x$$>$ for bar-crossing slices. The best-fit slope determination for $\Omega_p$ from such a plot is primarily governed by slices with the largest $<$$v$$>$ and $<$$x$$>$.  Since in SB0 galaxies
$<$$v$$>$ and $<$$x$$>$ approach zero in slices at or near the projected bar end (since they are presumably too far past the strong bar structure to contain 
information about the bar and mark, rather, a return to axisymmetry in the disk), even when TW estimates consider these `near zero' slices, they contribute 
minimally to estimates for $\Omega_p$.  (See \citealt{debPA} where studies with a simulated SB0 include a number of slices sampling the full extent of the bar.)  But when there is considerable asymmetric structure present beyond the end of the bar, similarly positioned 
slices will reflect this contribution, impairing measurement of the true bar pattern speed.

\begin{figure}
\begin{center}
\epsscale{.90}
\plotone{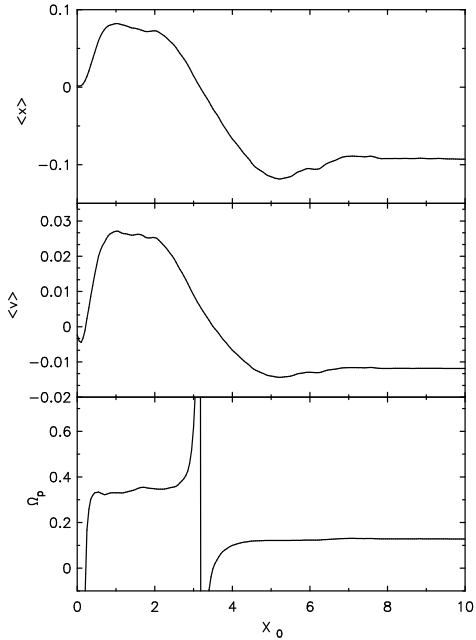}
\end{center}
    \caption{Variation of $<$$x$$>$ (top), $<$$v$$>$ (center) and $\Omega_p^{TW}$=$<$$v$$>$/$<$$x$$>$ (bottom) with $X_0$ for the 
slice at $y$=1.2 with SA=15$\degree$ for the barred spiral simulation.   
\label{fig-integrals75b}}
\end{figure}
In the barred spiral simulation, this consequence can be characterized upon inspection of $<$$x$$>$, $<$$v$$>$ and 
$\Omega_p^{TW}$=$<$$v$$>$/$<$$x$$>$ as a function of the limit of integration $X_{max}$=$X_0$ in a typical slice at $y$=1.2 (see Figure \ref{fig-integrals75b} for SA=-75$\degree$).  $\Omega_p^{TW}$ for this slice, which according to our estimate for $a_B$ 
crosses the outer region of the bar, actually seems more reflective of the spiral pattern speed (beyond the discontinuity at 
$X_0$$\sim$3.2).  Indeed, past $X_0$$\sim$3.2 where $<$$x$$>$ crosses zero we can infer that this bar-crossing slice 
contains substantial participation from the spiral; both $<$$x$$>$ and $<$$v$$>$ decrease to a dip between $X_0$=4.0 and 
6.0 before reaching a plateau.  This is evidence that, despite the relative weakness 
and limited extent of the spiral as indicated by the surface brightness distribution, the asymmetry in the disk should 
not be considered dominated by the bar alone.  Moreover, this is a clear indication that 
the TWR calculation, which identifies and effectively removes the spiral contribution from the fully extended 
$\Sigma$-weighted velocity integrals, can improve upon traditional TW bar estimates that also use fully extended 
integrals. Of course, as displayed by Figure \ref{fig-integrals75b}, integrating only between $\pm X_0$$\sim$ 2.5 
may alone provide a reasonable bar pattern speed and relieve all other TW slices from the spiral contribution.  
However, this would provide only a single pattern speed estimate where two are possible; using the TWR method and 
fully extended integrals, both $\Omega_b$ \textit{and} $\Omega_s$ can be measured.

Figure \ref{fig-integrals75b} also raises a crucial point related to the required limit of integration along 
each slice in the TWR calculation: if there is a clear plateau in the integrals reached before the edge of the map 
boundary, why not simply truncate the integrals where they have converged (common to TW estimates) rather than use 
integrals extending to the edge of the surface brightness and which present outermost bins that we demand must be 
`cut' (that is, calculated without regularization), anyway?  The plateau reached at $X_0$$\sim$8.0 would suggest that 
truncating the integral there could suitably account for information from the major sources of asymmetry in the disk.  
However, this same distinction is not clearly shared by all slices, especially those at large $\vert y\vert$.  If we associate 
the plateau in this slice at $y$=1.2 with the limit of integration $X_{max}$=8.2 and hence the total radial extent 
$R_{max}$ required of the quadrature where $R_{max}=\sqrt{X_{max}^2+(y/\cos{\alpha})^2}$, 
then this locates the outermost slice position at $y_{max}$=$(R_{max}-\Delta r)/\cos{\alpha}$ as well as the extent of the integral along this slice.  It is easy to check that (at least for this slice orientation) the integral has 
not achieved convergence by this point, nor have most other integrals in the disk by $R_{max}$.

Rather than risk ignoring information critical for characterizing the patterns uniformly throughout the disk, then, we choose to include in the quadrature all information out to the edge of the surface density.  Indeed, this serves to perform a function similar to truncating TW integrals.  The difference in the two procedures arises from the fact that while the limits of integration for each individual slice can be adjusted for a given structure of interest in the TW calculation, the quadrature in the TWR method delineates specific bounds which must encompass complete information from all extended patterns in the disk. 
\subsubsection{\label{sec:Err}Systematic Errors}
In this section we use the barred spiral simulation to consider the errors introduced to TWR pattern speed estimates in real galaxies.  We can expect errors in the assumed $PA_{disc}$ to dominate errors in the TWR 
calculation, given that such errors translate significantly to inaccuracies in the traditional TW calculation via the 
line-of-sight velocity integral which is also, of course, a prominent feature of the radial TW equation.  
Table \ref{tab-PAerror} summarizes the results for a standard PA error of $\delta_{PA}=\pm2\degree$ on the SAs 
in quadrant I chosen for their advantages, as evidenced by the discussion in $\S$ \ref{sec:iError}.  The average and rms for bar and 
spiral estimates from both the TWR and the traditional TW methods are listed.  (The TW `bar' estimates are obtained by fitting to the inner, nominal number of slices listed in Table \ref{tab-BStw2}, while all slices with $\vert y\vert$$<$5.1 are considered in the `spiral' estimates.)

Even this small 
$\delta_{PA}$ introduces considerable errors (relative to the known pattern speeds) to both types of bar estimates. These errors in $\Omega_b$ can be many times larger than the formal rms.  But whereas the errors are comparable in the TWR and TW bar estimates, the errors in the spiral estimates tend to be smaller with TWR than TW. Rewardingly, with this error
not only are the TWR spiral solutions still definitively constant and accurate to $\sim$15\%, but the radial domain of both pattern speeds are still well-determined. The transition between the bar and spiral $r_t$ and the termination radius of the strong spiral pattern $r_c$ are effectively unchanged from the $\delta_{PA}=0$ case; for both $\delta_{PA}=+2\degree$ and $\delta_{PA}=-2\degree$ we find $r_t$=2.8$\pm$0.37and $r_c$=5.4$\pm$0.46.
\begin{table}
\begin{center}
\caption{PA errors in TW and TWR estimates for Simulation I. 
\label{tab-PAerror}}
\begin{tabular}{rcccc}
\\
\tableline\tableline
\multicolumn{1}{c}{} &\multicolumn{2}{c}{$\delta_{PA}=+2\degree$}&\multicolumn{2}{c}{$\delta_{PA}=-2\degree$}\\
\tableline
 &\textbf{$\Omega_b$}&\textbf{$\Omega_s$}&\textbf{$\Omega_b$}&\textbf{$\Omega_s$}\\
\tableline
TW*&0.179&0.162&0.354&0.115\\
  &$\pm$0.055&$\pm$0.058&$\pm$0.04&$\pm$0.02\\
TWR&0.214&0.211&0.380&0.158\\
  &$\pm$0.031&$\pm$0.045&$\pm$0.042&$\pm$0.023\\
\tableline
\end{tabular}
\tablecomments{Entries correspond to average bar and spiral estimates from SAs in quadrant I (-15$\degree$,-45$\degree$, and
-75$\degree$) with PA errors $\delta_{PA}=+2\degree$ and $-2\degree$.  Estimates from both the traditional TW (*bar estimates using the nominal number of slices for each SA listed in Table \ref{tab-BStw2} and TWR methods are listed).}
\end{center}
\end{table} 

Besides the effects on $PA_{disc}$ measurement as studied by \citet{debPA}, galaxy inclination and 
ellipticity play perhaps more prominent roles as sources of error in TWR solutions relative to traditional TW estimates. Presumably, large inclination 
errors will prevent the association of information into accurate radial bins, given that $r=y_{obs}/\cos \alpha  $. We expect this 
effect to be minimal at moderate inclinations since $dr\propto d\alpha \sin \alpha$, and most significant at small inclinations where 
one would generally find that the difficulty in inferring in-plane morphology and kinematics makes
the the TW method impractical in any case. At a moderate, 45$\degree$ inclination we find that the barred spiral solutions for $\pm$ 3$\degree$ 
inclination error differ from the actual pattern speeds by only the change in $\sin \alpha$ introduced by the line-of-sight velocity.
\subsubsection{\label{sec:Trans}Transition Mis-identification}
We have shown that the regularized TWR method can be used to parameterize the number and radial domain of 
multiple pattern speeds in a single disk.  Formally, the contribution of each to the line-of-sight velocity integral is 
established through the designation of a transition between patterns.  In our scheme this transition is a free parameter, but the method, of course, could plausibly assimilate other transition-identification methods to similar effect, much like that in the \citet{mac06} adaptation of the TW method.  Specifically, in \citet{mac06} a plateau in the integrals 
$\int_{-X_0}^{X_0} \Sigma x dx$ with variation in $X_0$ is associated with the transition from 
an inner to an outer pattern (see \citet{mac06} for details).  This transition is then used to separate the disk surface brightness 
into two unique components (one for an inner secondary bar, one for an outer primary bar), thereby governing the decoupling of the pattern speeds.

\begin{figure}
\begin{center}
\epsscale{.90}
  \plotone{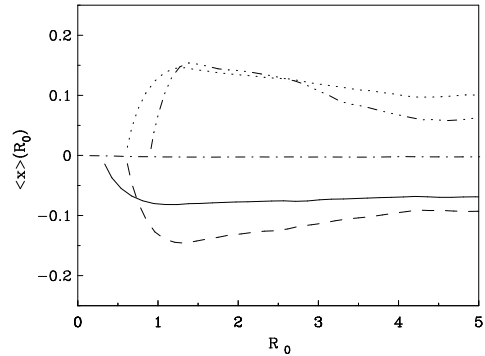}
\end{center}
    \caption{Variation of $<$$x$$>$=$\int_{-X_0}^{X_0} \Sigma x dx$/ $\int_{-X_0}^{X_0} \Sigma dx$ with $R_0$=$\sqrt{y_{i}^2+X_0^2}$ for 
five slices at SA=45$\degree$ ($y_1$=0.0 (dash-dot); $y_2$=-0.6 (dash); $y_3$=-0.3 (solid); 
$y_4$=0.6 (dot); $y_5$=0.9 (dash-dot-dot)) in the barred spiral simulation.  The plateau range indicates 
a transition between $R_0\sim$1.0-1.5.
\label{fig-MacInt}}
\end{figure}

We here pursue this type of diagnostic for the case of the simulated barred spiral in order to test the reliability of employing the TWR method with such independent evidence for pattern extent.  Figure \ref{fig-MacInt} plots the values of $\int_{-X_0}^{X_0} \Sigma x dx$ / $\int_{-X_0}^{X_0} \Sigma dx$  as a 
function of $R_0$ for five bar-crossing slices ($y_1$=0.0, $y_2$=-0.6, $y_3$=-0.3, $y_4$=0.6, $y_5$=0.9) at 
SA=45$\degree$ where $R_0$=$\sqrt{y_{i}^2+X_0^2}$.

\begin{table}
\begin{center}
\caption{TWR estimates for Simulation I with a mis-identified transition.\label{tab-BSMac}}
\begin{tabular}{rcccc}
\\
\tableline\tableline
SA&\textbf{$\Omega_b$}&\textbf{$\Omega_s$}&$r_t$&$r_c$\\
\tableline
75$\degree$&0.376&0.193&0.9&5.7\\
45$\degree$&0.348&0.152&0.9&4.8\\
15$\degree$&0.367&0.191&1.2&4.5\\
-15$\degree$&0.336&0.186&1.2&4.5\\
-45$\degree$&0.336&0.212&1.2&5.7\\
-75$\degree$&0.313&0.200&1.2&5.7\\
\tableline
\end{tabular}
\tablecomments{All bar and spiral pattern speeds listed here are estimated from TWR solutions where $r_t$ is restricted to between 0.9 and 1.5.  As in Table \ref{tab-BSall}, solutions are calculated using a $\Delta r$=0.3 bin width.  The third and fourth 
columns list the connate estimates for $r_t$ and $r_c$.}
\end{center}
\end{table} 
For this projection, there seems to be a plateau at $R_0\sim 1.2$. We note that this value is 
smaller than 
the bar pattern extent indicated by our best-fit solutions (and the major axis bar length estimated by inspection of 
the surface brightness distribution) and furthermore, the same analysis performed 
over the range of SAs does not always as clearly show the same behavior.  Presumably, this particular value is more 
indicative of the bar minor axis length than the full radial zone of the bar pattern, since slices at a 45$\degree$ SA sample 
along $X$ perpendicular to the bar major axis; other slice orientations are similarly limited to sampling the 
bar according to its specific projection.

Allowing that information from the full radial zone of the bar is not manifest in this type 
of indicator, we here proceed to assess the consequences for inner and outer pattern speed estimates when the transition between the two patterns is misidentified.  Table \ref{tab-BSMac} lists 
the TWR bar and spiral estimates for each of the SAs studied in the previous sections where we have limited the transition to  0.9$<$$r_t$$<$1.5.  At all SAs, this error of several bins in $r_t$ causes an overestimation of the inner pattern speed.  The outer pattern speed, on the other hand, though slightly raised, is still reassuringly accurate.  We can interpret the inaccuracy in $\Omega_b$, then, as the result of the mis-association of information from one pattern to the other via equation (\ref{eq:outin}), as discussed at the end of $\S$ \ref{sec:iError}.  Additionally, we can attribute the greater inaccuracy in solutions from quadrant II to the reasons discussed in $\S$ \ref{sec:iError}.  Though the subtleties in Table \ref{tab-BSMac} are most likely specific to this simulation, we emphasize that systematic pattern speed errors introduced by transition mis-identification are generic to the nature of the calculation.

One of the greatest strengths of the TWR calculation is that the transition is in principle a free parameter (within limits) 
and need not be restricted to a single, pre-determined value.  We therefore recommend letting the results of the TWR calculation speak for themselves: given sufficient resolution and reasonable measurement errors, step-model solutions with the most realistic transition should be recognizable by how well they reproduce the actual $<$$v$$>$. Since the transition determines the separation of the patterns by interpreting the contribution made to these integrals by each, the natural result is the most accurate determination of the pattern speeds possible.

\subsection{\label{sec:SIM2}Simulation II: Spiral Galaxy}
\begin{figure}
\begin{center}
\epsscale{.90}
%  \plotone{file=CUTspiral.eps, height=6cm,width=6cm
   \plotone{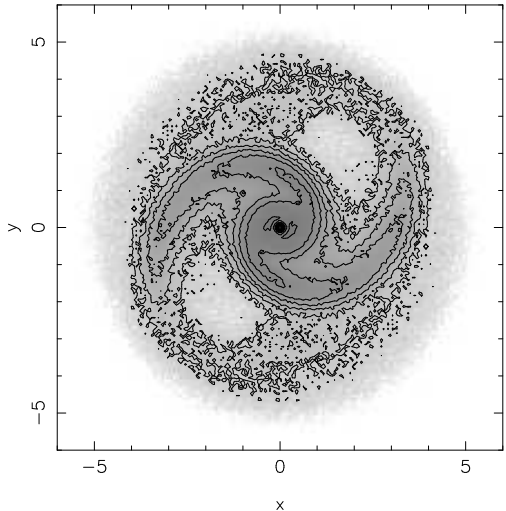}
\end{center}
    \caption{Face-on display of the spiral simulation's surface brightness distribution projected with a -30$\degree$ rotation about the $z$ axis, with orientation as in Figure \ref{fig-SIMI}.  
\label{fig-SIM2}}
\end{figure}
\begin{figure}
\epsscale{.90}
\begin{center}
   \plotone{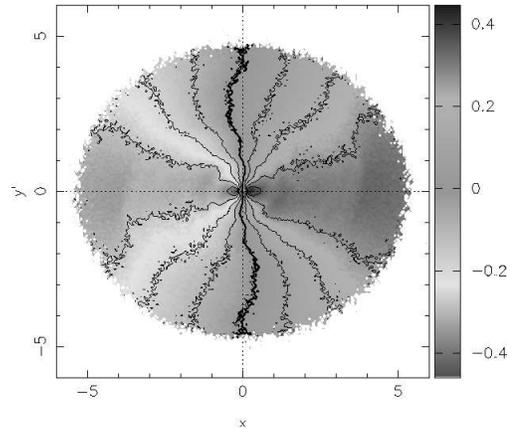}
\end{center}
    \caption{Face-on display of the spiral simulation's velocity field projected with no rotation about the $z$ axis shown here with the kinematical major axis running from left to right.  Contours are spaced at $\Delta v$=0.1.
\label{fig-SIM2vel}}
\end{figure}
In the previous section we showed that the TWR method is capable of detecting and measuring a constant spiral pattern 
speed that spans less than a third of the disk.  Here, we test the aptitude of the TWR method in measuring a 
radially varying spiral pattern speed that subsists over a large 
radial zone.  Since the strong spiral surface density enhancement in this simulation (Figure \ref{fig-SIM2}) has 
only moderate azimuthal range like the spiral in $\S$ \ref{sec:SIM1}, we further explore the likely limitations 
intrinsic to detecting spiral nonaxisymmetry with a given slice orientation.

\begin{figure}
\epsscale{.90}
\begin{center}
  \plotone{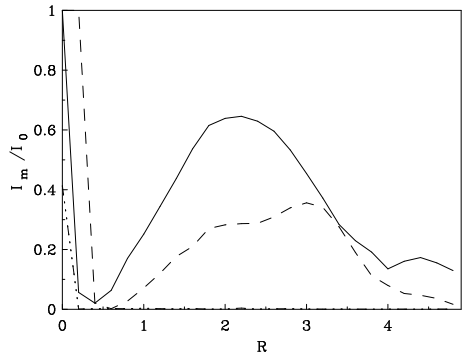}
\end{center}
    \caption{Fourier power spectrum of the spiral simulation's surface brightness distribution shown in Figure \ref{fig-SIM2}.  Modes up 
to $m=4$ are plotted as a function of radius with lines for $m=1$ in dot, $m=2$ in solid, $m=3$ in dash-dot-dot, and $m=4$ in dash.  \label{fig-pplot30S}}
\end{figure}

The two-armed spiral featured in this simulation extends over a large portion of the disk and is strong both in the surface 
brightness distribution (Figure \ref{fig-SIM2}) and as traced by departures from axisymmetric rotation (streaming 
motions) in the velocity field (Figure \ref{fig-SIM2vel}).  We estimate the extent of the spiral structure from that of 
the dominant $m$=2 component in the Fourier power 
spectrum plotted in Figure \ref{fig-pplot30S}.  With the expectation, then, that the spiral structure exists between 
0.5$\lesssim$$r$$\lesssim$3.5, we restrict our spiral pattern speed solutions to the radial zone 
$r_t$$<$$r$$<$$r_c$ bordered at the innermost and outermost radii by two independent sets of unregularized bins.  (For our models, 0$<$$r_t$$<$1.0 and 2.8$<$$r_c$$<$4.2.)  Between the radii marked 
by the free (though restricted) parameters $r_t$ and $r_c$, we allow each spiral pattern speed solution to vary with radius as an $n^{th}$ order polynomial where $n$=0 to 2.

The two sections of `place-holding' unregularized bins serve to isolate the solution in the radial range of 
interest; by minimizing errors due to either incorrectly incorporating or imposing an ill-prescribed form to bins outside the radial 
zone of the spiral, we reduce the introduction of inaccuracies to the spiral pattern speed solution.  The first, 
outermost section covers the radial zone identified by visual 
inspection of the surface brightness (and substantiated by the Fourier power spectrum) where the dominant spiral ceases 
to extend and where the weakness of the Fourier components suggests a region without a noticeable pattern.  As a result, we consider this zone unsuitable for reliable extraction of a pattern speed. (This last point can also be evidenced by traditional TW values from 
slices that pass solely through the outermost disk; the 
low counts there lead to a large degree of variation in $\Omega_p$ estimates from slice 
to slice).   The second, innermost section 
covers the radial zone inside $r_t$$\sim$1.0.  Within this radius we do not expect to be able to extract a realistic 
pattern speed solution since neither the morphology, velocity field, nor the power spectrum indicate a departure from axisymmetry.

Lacking such evidence in the innermost radii, it would be entirely 
possible to proceed without the use of unregularized inner 
bins (as would be the case when the 
indicators for such a measure are perhaps less obvious).  Indeed, allowing an independent regularized solution to exist at 
$r$$<$$r_t$, the TWR calculation still performs well; since these bins cover a rather small region of the 
disk and thereby contribute minimally to the $\chi^2$ through only the innermost slices, we find that the spiral solution outside of 
this region closely resembles that in the case where the inner bins are unregularized.  However, we proceed in 
the manner described above with the expectation that, if only minorly, our spiral pattern speed solution will be 
improved.

\begin{figure}
\epsscale{.90}
\begin{center}
  \plotone{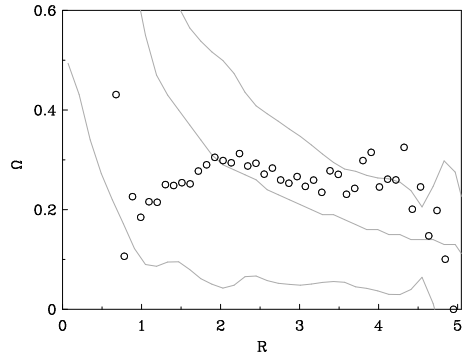}
\end{center}
    \caption{Plot of $\Omega_p$ as a function of radius for the spiral simulation as derived from the time evolution of the $m=$ 2 component.  Curves for $\Omega$ and $\Omega\pm\kappa$/2 are shown in gray.  
\label{fig-phaseS}}
\end{figure}

The actual pattern speed at the time of the snapshot shown in Figure \ref{fig-SIM2} as derived from the time 
evolution of the phase of the $m$=2 component is plotted as a function of radius in Figure \ref{fig-phaseS}.  This 
plot confirms that within $r$$\sim$1.0, the pattern speed is ill-quantified, with the values for $\Omega_p$ at 
the inner-most radii oscillating between positive and negative values outside of the vertical range of the plot.  At 
the largest radii, the pattern speed is characterized by scatter presumably reflective of the lack of a noticeable pattern in Figure \ref{fig-SIM2}.

The pattern speed between 1.0$\lesssim$$r$$\lesssim$3.0 to be reproduced by our solutions shows 
high-order variation with radius. Inside of 
$r$$\sim$2.0 where the pattern speed is at a maximum ($\Omega_{p,max}$$\sim$0.3), the pattern seems to 
be unwinding, while at larger radii the pattern speed decreases with increasing radius.

\begin{figure*}
\begin{center}
\epsscale{.90}
\plottwo{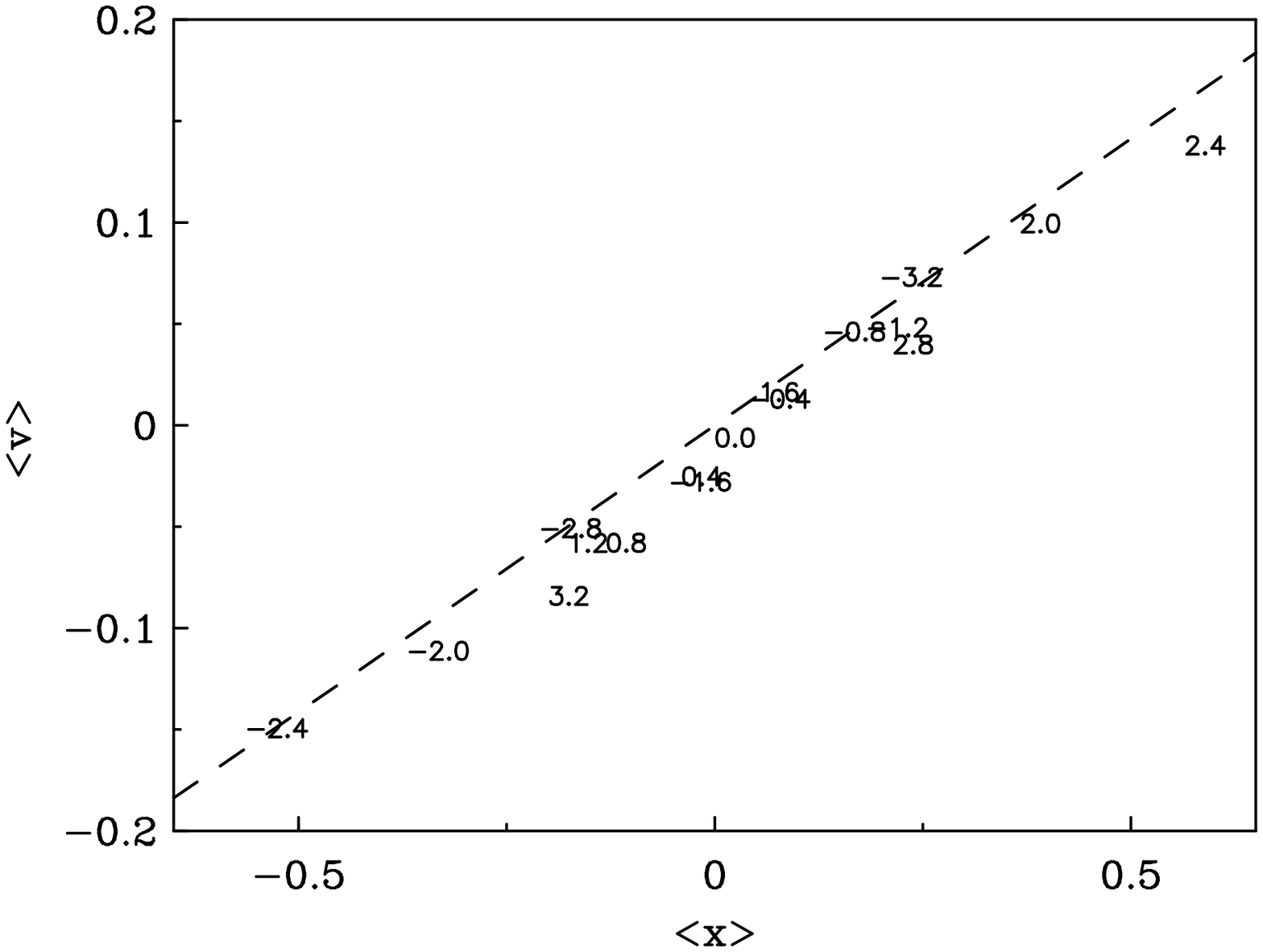}{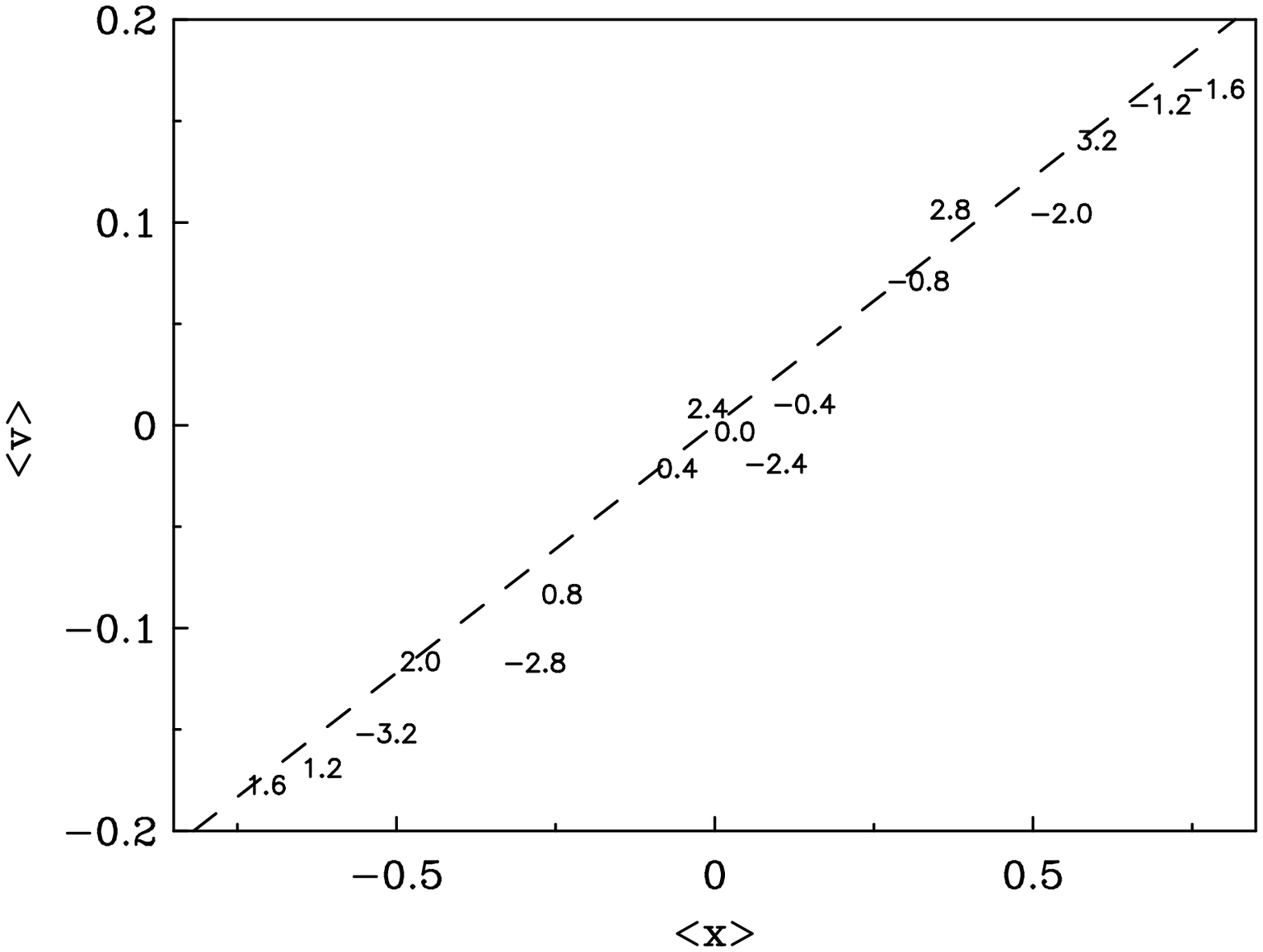} \\
\plottwo{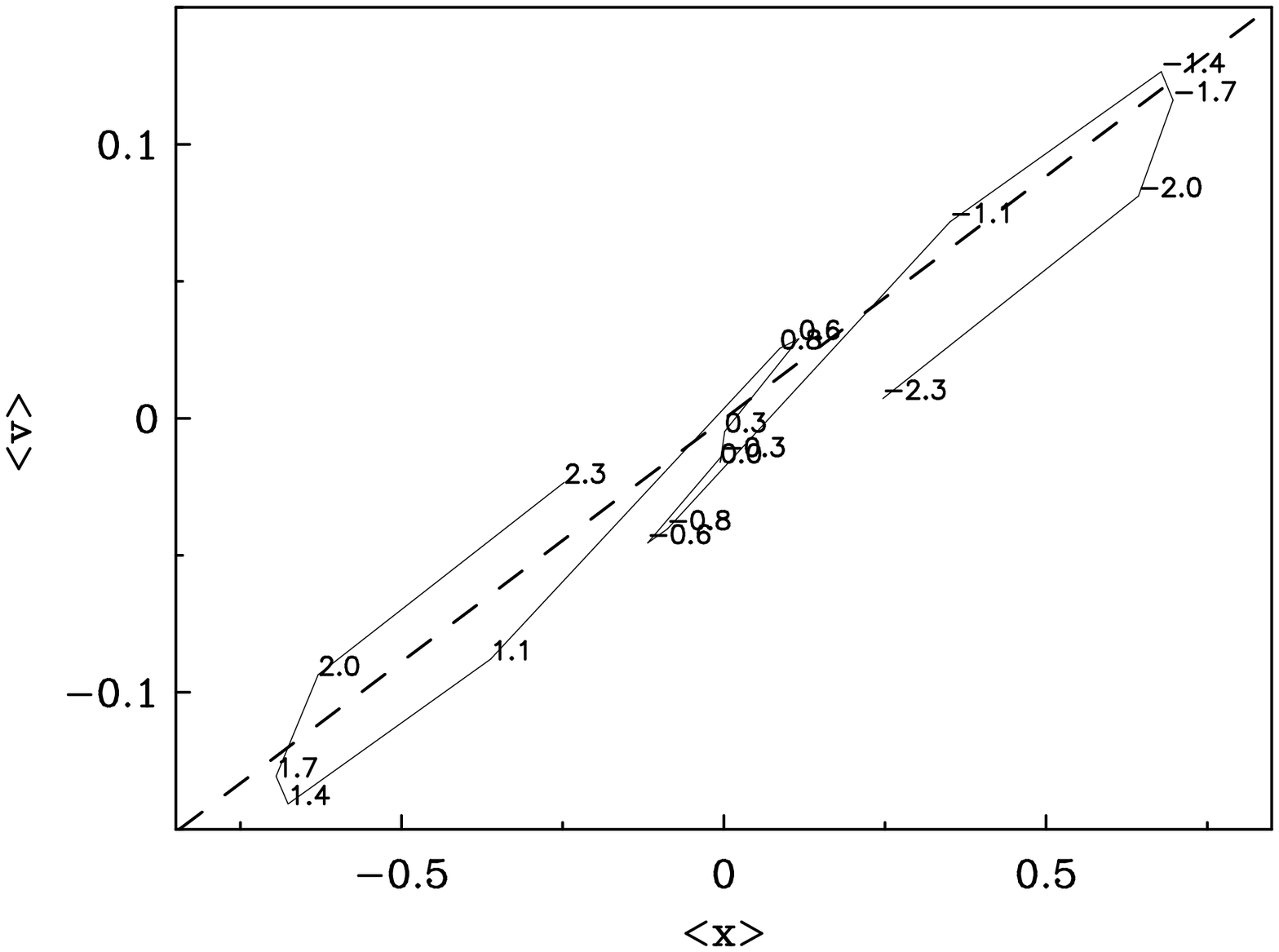}{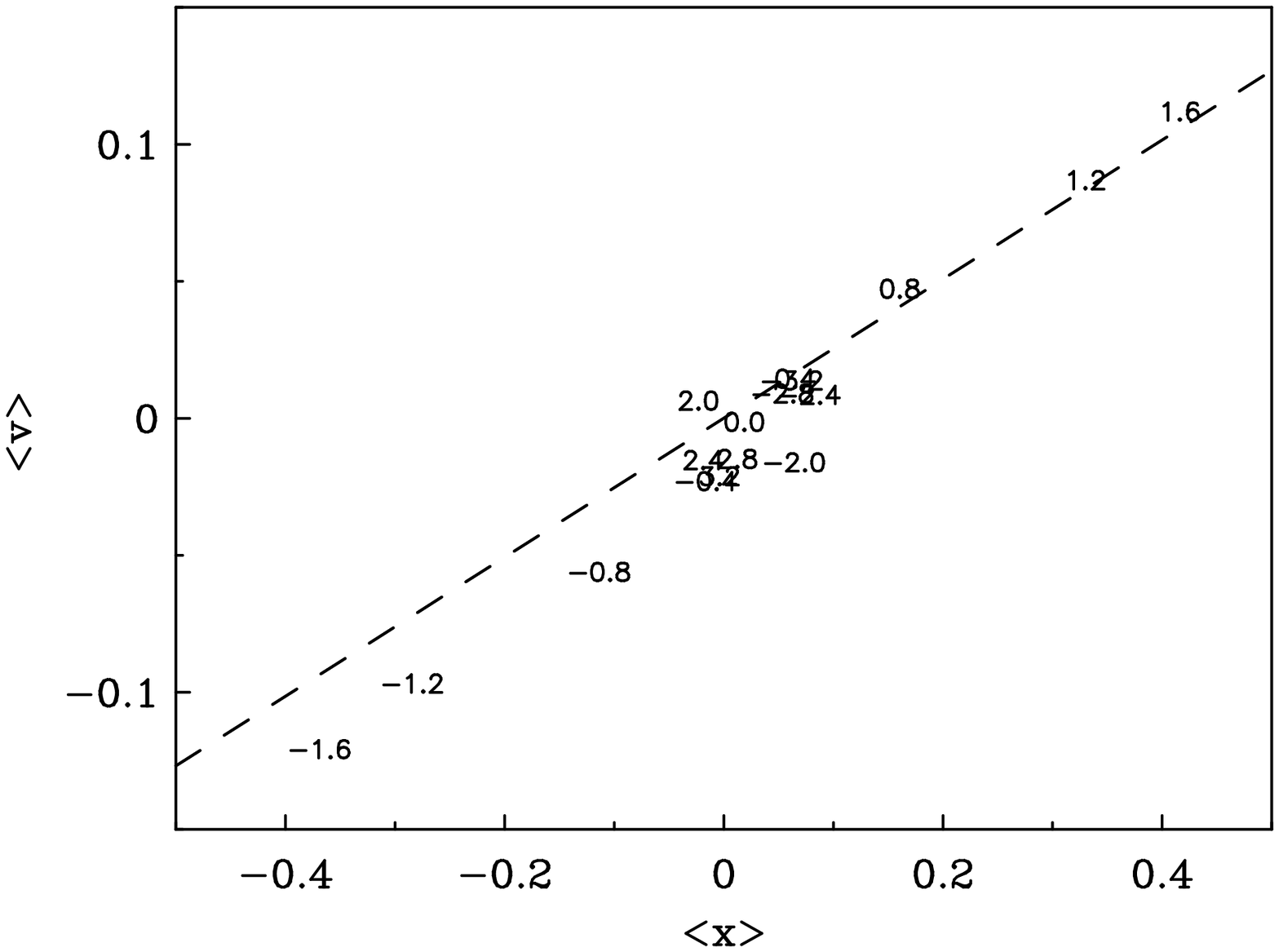} 
\end{center}
\caption{Plot of $<$$v$$>$ vs. $<$$x$$>$ for every other slice spaced at $\Delta y$=0.4 for the spiral simulation at four SAs (clockwise from top 
left: 45$\degree$, 15$\degree$, -15$\degree$, and -45$\degree$).  Each slice is labeled by its distance 
from the galaxy major axis $y_i$.  Adjacent apertures in the -45$\degree$ case are connected by a solid 
line; this SA shows the clearest signature of winding.  The dashed line in all plots is the best-fit straight line to all apertures shown. \label{fig-tw51}}

\begin{center}
\epsscale{.90}
\plottwo{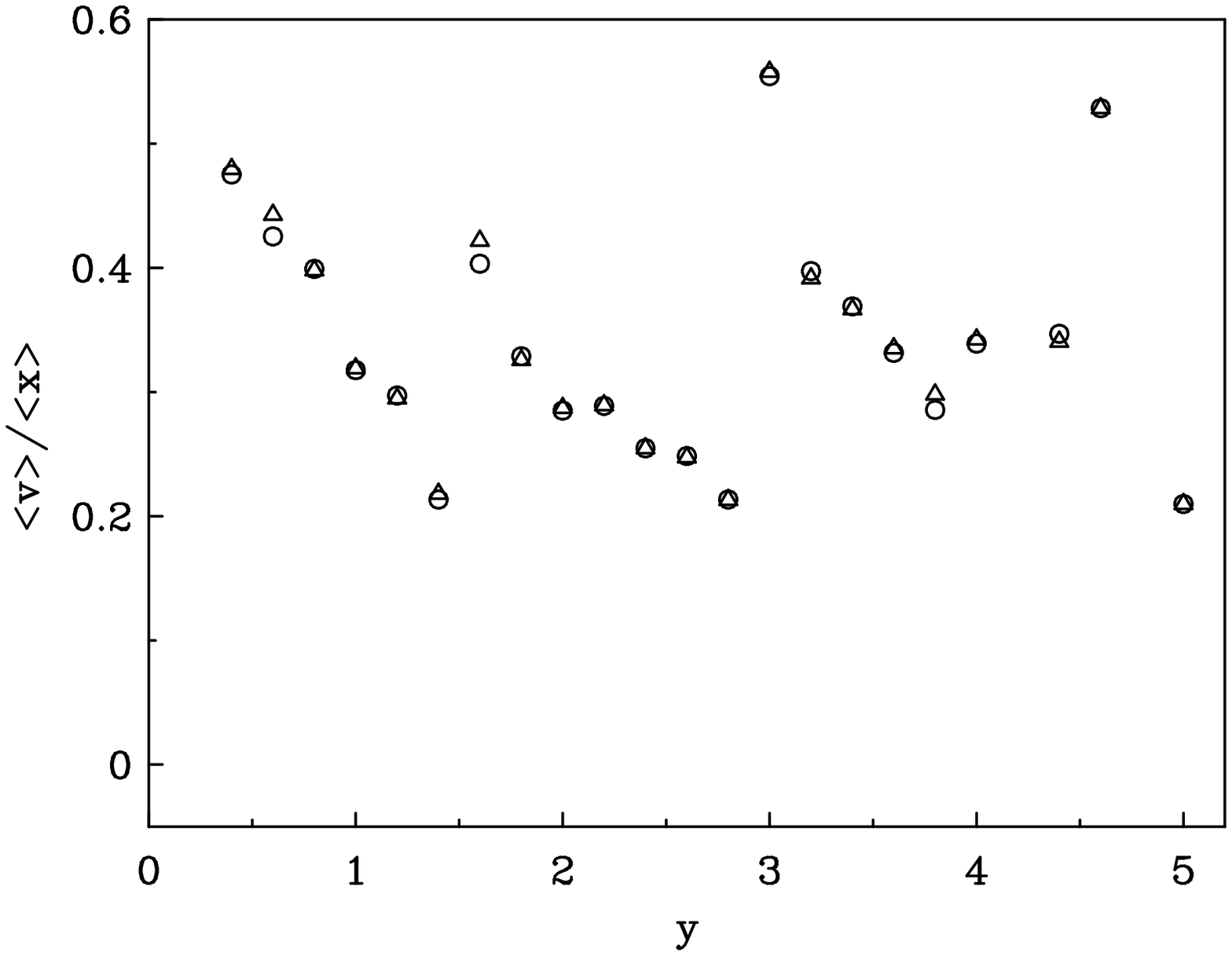}{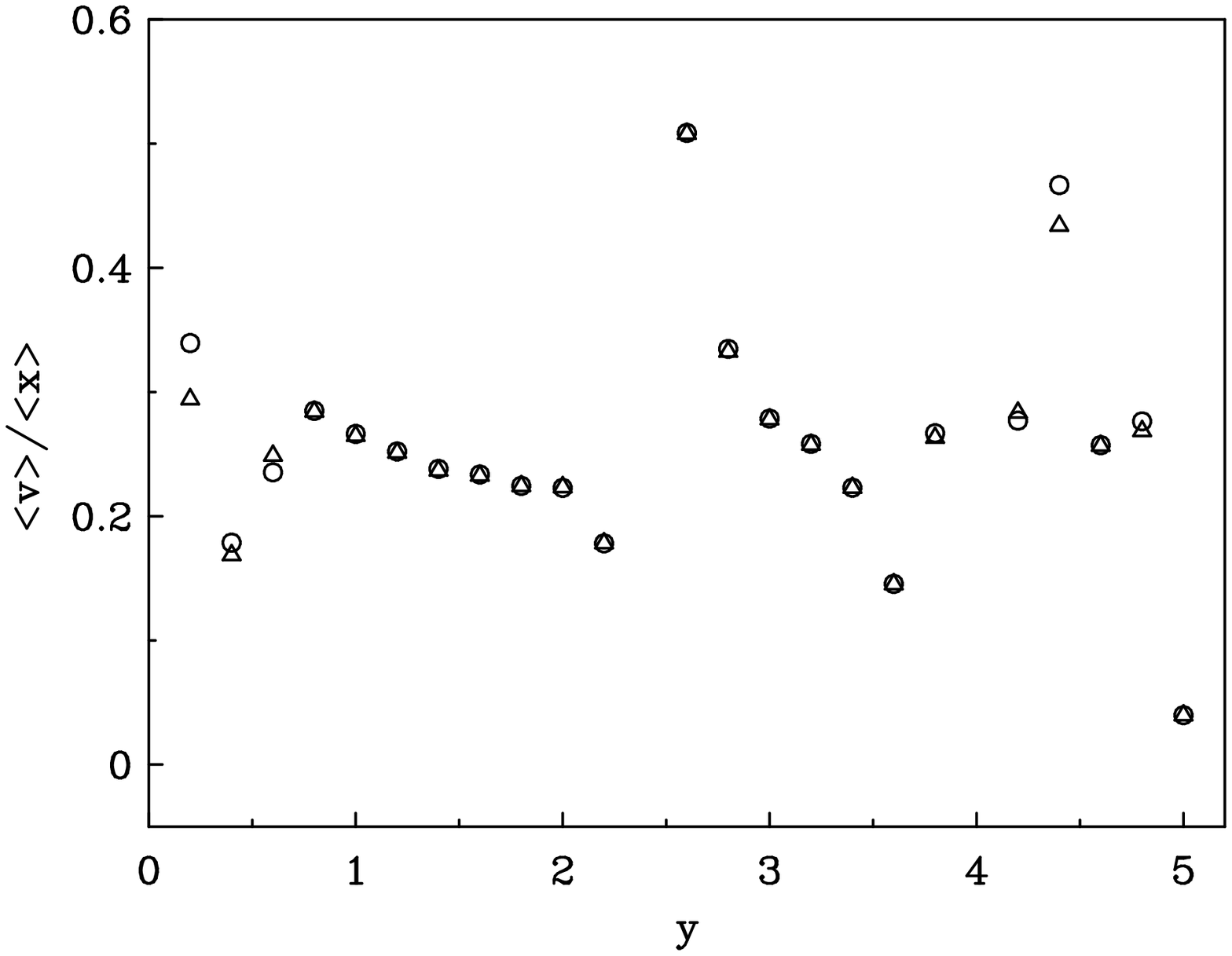} \\
\plottwo{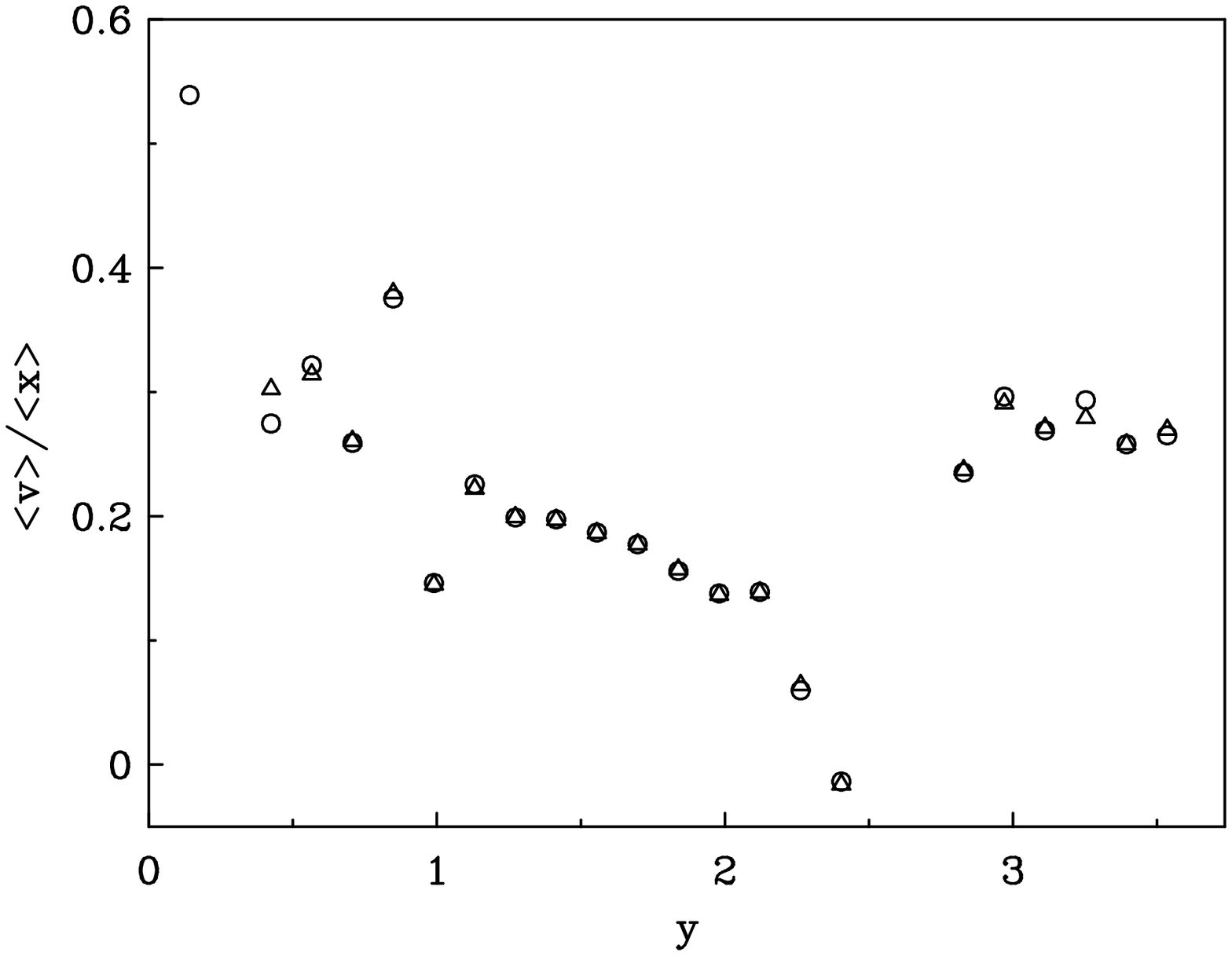}{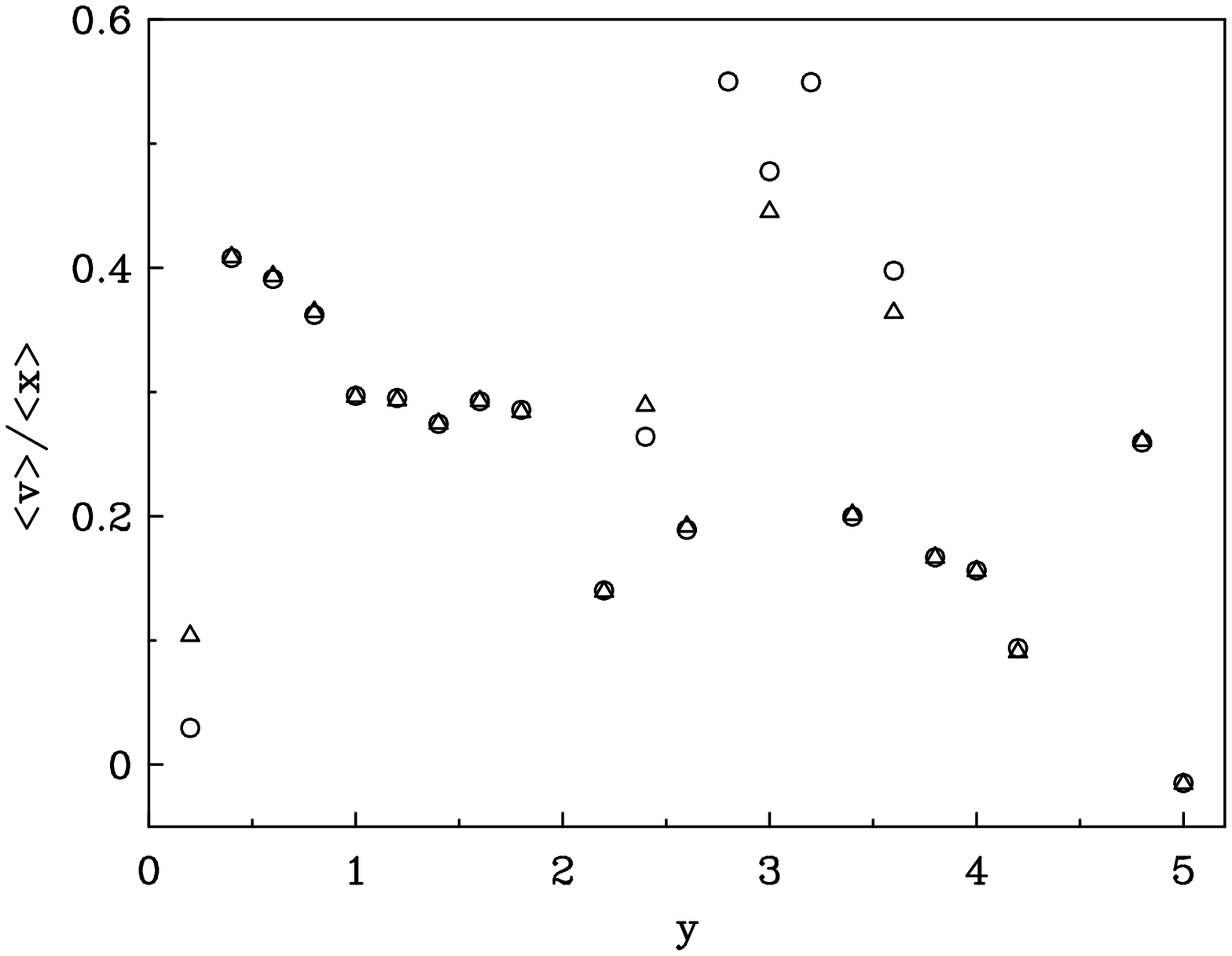} 
\end{center}
\caption{Plot of $\Omega_p$=$<$$v$$>$/$<$$x$$>$ as a function of distance from the galaxy major axis $y$ for the spiral simulation at the four 
SAs in Figure \ref{fig-tw51} (clockwise from top left: 45$\degree$, 15$\degree$, -45$\degree$, and -15$\degree$).  Triangles (circles) mark slices on the y$>$0 (y$<$0) side of the galaxy.\label{fig-twslice51}
}
\end{figure*}
This behavior is only modestly indicated by traditional TW estimates.  In plots of $<$$v$$>$ vs. $<$$x$$>$ 
(see Figure \ref{fig-tw51} for a comparison of plots generated at four SAs), 
adjacent slices trace out a figure-of-eight shape characteristic of complex radial behavior (namely winding).  However, 
the evidence for a variable best-fit slope (expected for a pattern speed that 
unwinds and winds) is not comprehensive or even readily apparent in all cases. (In fact, all slices are seemingly well fit with a single slope.)  The radial dependence of $\Omega_p$ becomes more apparent upon inspection of the variation of 
$\Omega_p$ with slice position (see Figure \ref{fig-twslice51}), but again, the radial dependence gets 
smoothed out since $\Omega_p$ for each slice is the result of averaging over all sampled radii.  Furthermore, for the 
four SAs shown in Figure \ref{fig-twslice51}, there is no single radial behavior implied by all.

\begin{figure}
\epsscale{.90}
\begin{center}
  \plotone{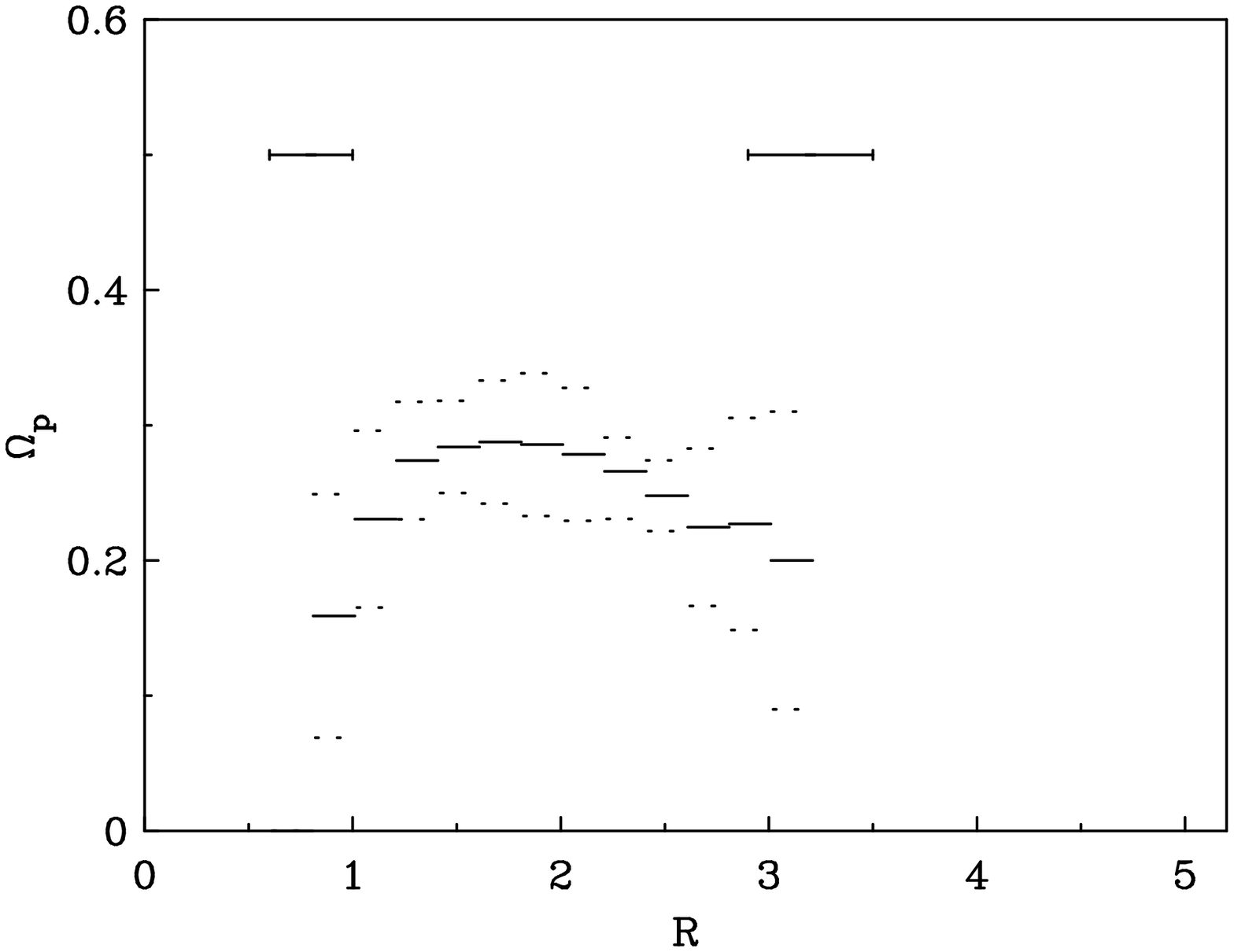}
\end{center}
    \caption{The best-fit regularized solution and error bars from the TWR method using $\Delta r$=0.2 bins as applied to the spiral simulation 
averaged over six SAs.  The best solution is 
shown as a series of solid lines for each bin with dashed errors. Errors for the transitions $r_t$=0.8$\pm$0.2 and $r_c$=3.2$\pm$0.3 are represented by horizontal error bars at the top.\label{fig-Ssol}}
\end{figure}
The TWR solutions, on the other hand, are capable of clearly displaying high-order variation in $\Omega_p(r)$.
The average and rms of the best-fit solutions with $\Delta r$=0.2 bins at six different SAs ($\pm$15$\degree$, $\pm$45$\degree$, and 
$\pm$75$\degree$) is plotted in Figure \ref{fig-Ssol} between the average values of $r_t$ and $r_c$.  For each SA, we find that the best-fit solution is 
quadratically varying, correctly reproducing the winding behavior present in the actual 
pattern speed.  However, the parameters $r_t$, $r_c$ and the size and location of $\Omega_{p,max}$ differ slightly 
in each solution.  These differences tend to reflect the influence of morphology and slice orientation, as in $\S$ \ref{sec:iError}.  The value of the peak in $\Omega_p(r)$, for example, is higher and more pronounced in solutions from quadrant II than quadrant I; at positive (negative) SAs $\Omega_{p,max}$$\sim$0.34 (0.27), on average.  Together with differences in the location where $d\Omega_p(r)/dr=0$ (which varies for the 
six SAs within the range 1.6$<$$r$$<$2.2), upon averaging, the result is a slightly under-measured peak value occurring at $r$$\sim$1.8 (in accord with the actual location).

The vertical error bars on 
bins at large and small $r$ similarly reflect variations in the location of $r_c$ and $r_t$ with SA.  Principally, the solution suffers contamination from bins that perhaps yet contain information from the inner or outer axisymmetric zones.  But we also find that solutions in quadrant II tend to decrease from $\Omega_{p,max}$ to a value $r_c$ that is further in by about 10\% than for solutions using slices at the perpendicular orientation.  Nevertheless, we find the average $r_t$=0.8$\pm$0.2 and $r_c$=3.2$\pm$0.3 to be in agreement with the bounds of the spiral pattern indicated by the disk surface density and its Fourier decomposition.

Despite these PA-dependent effects, the values implied by the average solution 
($\Omega_{p,max}$, $\Omega_{p,min}\vert_{r=0.8}$, and $\Omega_{p,min}\vert_{r=3.2}$) are accurate to within 5\%, 11\% 
and 10\%, respectively.  That we have correctly reproduced the high-order variation of $\Omega_p(r)$, regardless of SA, however, is 
perhaps the most remarkable aspect of the TWR solutions, even though the detectable variation is only at the 30\% level.

Naturally, our solution for $\Omega_p(r)$ lends itself to pattern winding time estimates. With the average values for 
the maximum and minima implied by our solution we can estimate the winding time of the pattern according to
\begin{equation}
\tau_{wind}=2\pi/(\Omega_{p,max}-\Omega_{p,min}).
\end{equation}
For the outer spiral arm, for example, we estimate an average time to wind $\bar{\tau}_{wind}$= 71.64 which is less 
than 10\% from the actual winding time $\tau_{wind}$=78.53 observed from the time evolution of the simulation.  (This, of course, assumes that $\Omega_p(r)$ does not vary 
over this time.)

As this simulation would indicate, even without 
uniform slice coverage, though it may be slightly more difficult to determine with confidence the radial domain of the 
pattern (given large errors in $r_t$ and $r_c$), the overall shape, or functional form, for $\Omega_p(r)$ can be 
ascertained.  Of course, 
this is largely influenced by the adopted measurement errors $\sigma^{<v>}$ for each slice and the quality of a priori 
information that can be gathered and employed.  With larger errors $\sigma^{<v>}$, for instance, the $\chi^2$ criterion 
becomes less discriminating, and it may be difficult to distinguish between several different radial dependences for 
$\Omega_p(r)$.  Additionally, without clear evidence that limits where the spiral pattern terminates, we risk 
misidentifying intrinsic radial variation.  Indeed, if we restrict $r_c$ to less than that implied by 
the best-fit solutions--and search instead for solutions at a second $r_c$-$\chi^2 $ minimum--the pattern speed solutions for all 
six SAs are constant between $\bar{r}_t$$\sim$0.47 and 
$\bar{r}_c$$\sim$2.47.  As may be expected, the average value for this constant pattern speed $\Omega_p$=0.236$\pm$0.051 is similar to that suggested by traditional TW estimates where $\Omega_{p,TW}$=0.207$\pm$0.046 on average.  

\subsection{\label{sec:SIM3}Simulation III: Double Barred Galaxy}
\begin{figure}
\epsscale{.90}
\begin{center}
\plotone{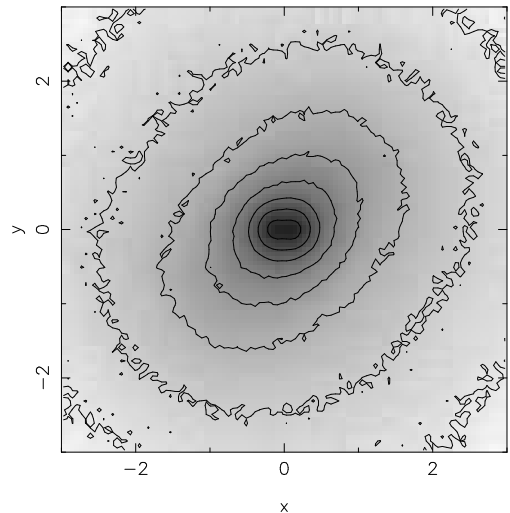}
%\plottwo{f18a.eps}{f18b.eps} 
\end{center}
\caption{Face-on display of the double barred simulation's surface brightness distribution projected without rotation about the $z$ axis, highlighting the inner two bars. \\
\label{fig-SIMIII}}
\end{figure}
\begin{figure}
\epsscale{.90}
\begin{center}
  \plotone{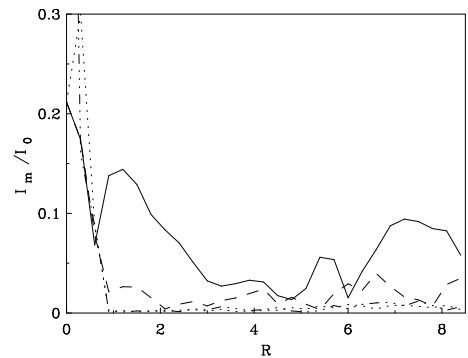}
\end{center}
    \caption{Fourier power spectrum of the double barred simulation's surface brightness distribution shown in Figure \ref{fig-SIMIII}.  Modes up 
to $m=4$ are plotted as a function of radius with lines for $m=1$ in dot, $m=2$ in solid, $m=3$ in dash-dot-dot, and $m=4$ in dash.\label{fig-2Bpower}}
\end{figure}
In this section we address the use of the TWR method for the purposes of nuclear bar detection and measurement using 
the double barred SB0 simulation pictured in Figure \ref{fig-SIMIII}.  In performing the regularized TWR calculation 
we again act under the assumption of multiple patterns in distinct radial zones.  Our models for $\Omega_p(r)$ 
parameterize unique, constant pattern speeds for both the primary and secondary bars, known to have pattern speeds of $\Omega_{pb}$=0.23 and $\Omega_{sb}$=0.41, respectively.  From inspection of the 
surface density and its Fourier decomposition (for which the power spectrum is 
plotted in Figure \ref{fig-2Bpower}), we associate the drop in power of the $m$=2 component at $r$$\sim$0.8 and again at 
$r$$\sim$3.0 with the end of each bar.  In step-models for $\Omega_p(r)$, then, we restrict the secondary-to-primary bar 
and primary bar-to-disk transitions to within 0.4$<$$r_{t,1}$$<$1.2 and 2.5$<$$r_{t,2}$$<$3.2, respectively.

To isolate the bars from the rest of the disk we extend the quadrature to the edge of the surface density
and employ our `cut' procedure in light of the argument set forth at the end of $\S$ \ref{sec:TWTWR}.  This is 
particularly compelling here since, despite the apparent axisymmetry beyond the primary bar in this SB0 simulation, 
the Fourier decomposition shows power in the $m$=2 mode beyond $r$$\sim$3.0, especially in the last third 
of the disk.  If the asymmetry in this radial zone (which appears only very weakly in the surface density) is sustained by 
an ill-defined, non-unique or perhaps unrigid pattern--for which the TW/TWR assumptions 
break-down--then associating it with a measurable pattern speed will likely introduce regularization-induced bias to 
the solution for the interior patterns of interest.

Critically, asymmetry such as this may prevent clear integral convergence beyond the 
primary bar end.  So, too, can its presence in the integrals be expected to degrade the reliability of pattern speed estimates 
for the structures of interest.  Removing the influence of the information in this outer radial zone by calculating the 
bins there without regularization is our best chance for accurate pattern speed measurement.  Given that the departure from axisymmetry manifest by a small nuclear bar will be relatively minor 
compared with that of other patterns in the disk, this is especially relevant for accurate measurement of $\Omega_s$.  In this case, non-axisymmetry on a comparable scale may easily upset 
this structure's contribution to the integrals and, if prescribed an incorrect pattern speed model, may introduce consequences 
for the inner most bins in solutions.

Our procedure for this simulation, however, does not quite involve calculating without regularization all bins up to the patterns of interest (i.e. to the end of the primary bar).  As in all cases, but particularly here where we are compelled to `cut' approximately two thirds 
of the disk, leaving a large portion of the total bins unregularized 
may begin to reintroduce unamendable propagating noise.  To reduce this risk, careful attention has been paid to the
development and testing of models which prevent the destabilization of solutions.  (Indeed, the appropriate 
balance 
between noise and stability in models for $\Omega_p(r)$ must be explored on a galaxy by galaxy basis.)  Here, we find that the most stable and realistic models for $\Omega_p(r)$ are 
those which include a third, constant pattern past the end of the primary bar.  From this we might  
infer that, though weak and difficult to discern in Figure \ref{fig-SIMIII}, there exists a spiral 
pattern outside the primary bar, perhaps corresponding to the $m$=2 component beyond $r$$\sim$3.0 which remains clear, though modest out to $r$$\sim$5.0.  Indeed, we assume that the third minimum in the power of the $m$=2 component at this radius corresponds to the end of the spiral pattern, and moreover, since counts are low in the rest of the disk, that the bins beyond $r$$\sim$5.0 are best calculated without regularization.  We note, however, that we do not necessarily expect to measure a realistic 
pattern speed in this third radial zone.

Compiling this evidence for two bars and a possible spiral we search for the best-fit solutions 
parameterizing two constant pattern speeds (one for the primary bar, one for the inner secondary 
bar) out to the end of the primary bar, in addition to a third constant pattern speed $\Omega_s$ restricted to extend out to 4.3$<$$r_c$$<$5.3.  We find that solutions generated in this manner provide much more accurate estimates for the primary and secondary bars compared with 
solutions that are either regularized over the full extent of the surface density or unregularized up to the end of the 
primary bar.

For the purposes of further establishing favorable conditions for nuclear bar detection, 
we adopt a small bin width $\Delta r$=0.15 in the quadrature.  As described in $\S$ \ref{sec:iError} and $\S$ \ref{sec:Trans}, the mismatch by a fraction of a bin width or more between the actual transition and that to which the solution is 
confined can have consequences for both inner and outer TWR pattern speed estimates.  Real nuclear bars will need to be well resolved in order to accurately separate the contributions of the two bars.

The secondary bar in this simulation is known to be non-rigidly rotating; in \citet{sdeb07} and in \citet{dshen07}, the amplitudes and pattern speeds of secondary bars formed in purely collisionless $N$-body simulations through the introduction of a rotating pseudobulge oscillate as the bars rotate through the companions in which they are nested.  In \citet{sdeb07}, TW estimates of the secondary bar pattern speed measured using bulge-only kinematics (the bulge supports the nuclear bar alone and a primary bar contribution need not be accounted for in the TW integrals) are subject to marked errors consistent with an origin in non-rigid rotation.  These errors result in estimates of $\Omega_{sb}$ too high on one side of the galaxy and too low on the other, in accord with being a manifestation of the oscillations driving radial pulsations which contribute with different signs on the two sides of the galaxy.  Cancellation between measurements from both sides of the galaxy in global regularized solutions reduces the effect of the oscillations.  In the discussion to follow, our focus is on sources of error in the TWR calculation other than this intrinsic effect. 

\begin{table}
\begin{center}
\caption{TWR estimates for Simulation III.\label{tab-2Btwr}}
\begin{tabular}{rcccccc}
\\
\tableline\tableline
SA&\textbf{$\Omega_{sb}$}&\textbf{$\Omega_{pb}$}&\textbf{$\Omega_s$}&$r_{t,1}$&$r_{t,2}$&$r_c$\\
\tableline
75$\degree$&0.420&0.231&0.146&0.75&3.15&4.35\\
45$\degree$&0.548&0.278&0.178&0.75&3.15&4.35\\
15$\degree$&0.407&0.231&0.063&0.6&2.7&4.8\\
-15$\degree$&0.404&0.264&0.087&0.75&2.7&5.25\\
-45$\degree$&0.477&0.259&0.233&0.9&3.15&4.35\\
-75$\degree$&0.375&0.225&0.028&0.9&3.15&4.5\\
\tableline
-&0.41&0.23&-&0.8&3.0&5.0\\
\tableline
\end{tabular}
\tablecomments{The secondary bar, primary bar and spiral pattern speeds listed here are estimated from TWR solutions calculated using a $\Delta r$=0.15 bin width for a range of SAs.  The last three columns list the connate estimates for $r_{t,1}$, $r_{t,2}$ and $r_c$.  Values for the actual pattern speeds are shown in the last row.}
\end{center}
\end{table} 
\begin{figure}
\epsscale{.90}
\begin{center}
  \plotone{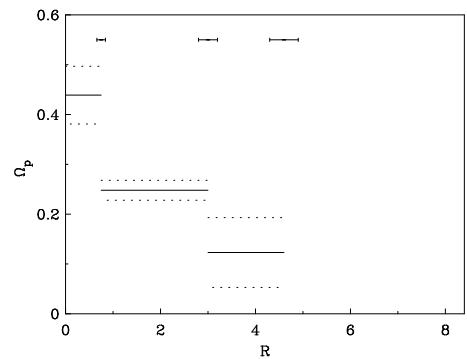}
 \end{center}
    \caption{The best-fit solution and error bars from the TWR method as applied to the double barred simulation averaged over six SAs using $\Delta r$=0.15 bins.  The secondary bar 
$\Omega_{sb}$=0.439$\pm$0.058, primary bar $\Omega_{pb}$=0.248$\pm$0.02, and spiral $\Omega_s$=0.123$\pm$0.07 are 
shown as solid lines with dashed errors. Horizontal error bars at the top indicate the dispersion in the secondary-to-primary bar 
transition $r_{t,1}$=0.75$\pm$0.09, the primary bar-to-spiral transition $r_{t,2}$=3.0$\pm$0.21, and spiral termination radius 
$r_c$=4.6$\pm$0.33.  
\label{fig-2BSsol3PAs}}
\end{figure}
The TWR estimates from solutions with $\Delta r$=0.15 bins at six SAs ($\pm$15$\degree$, $\pm$45$\degree$, and $\pm$75$\degree$) are listed in Table \ref{tab-2Btwr} and the average and rms for this SA range is shown in Figure \ref{fig-2BSsol3PAs}.  Compared with solutions generated using $\Delta r$=0.3 bins in the calculation, these solutions are much more stable and accurate; at less than 8\% from their actual values, the comprehensive primary and secondary bar pattern speeds evidently benefit from the integrity of the calculated transitions between patterns (we find $r_{t,1}$=0.75$\pm$0.09, $r_{t,2}$=3.0$\pm$0.21, and $r_c$=4.6$\pm$0.33).

Though the TWR estimates in Figure \ref{fig-2BSsol3PAs} are accurate, the rms in each suggests 8 to 13\% error on the part of the method.  In addition to the intrinsic non-rigid rotation-error expected for the secondary bar measurement, this may yet indicate the slight misdesignation of transitions between patterns, and we can also assume that (though we do not necessarily consider its measurement realistic) the third measured pattern speed has a non-trivial influence on the inner pattern speeds, as must $\Omega_{pb}$ on $\Omega_{sb}$, for instance.  

With so many free parameters in our model--more than in any other used in this paper--identifying unequivocally the dependence of each on the others is difficult (indeed, the rms in each estimate likely reflects the combined consequence of many such inter-dependencies).  As in previous sections, however, we can begin to understand the largest source of variability in the estimates from SA to SA by considering the influence of morphology and slice orientation on the solutions.  If we consider that the major axes of the two bars are oriented at $\sim$45$\degree$ from one another at this time step (that is, with the secondary bar aligned along the x-axis, the primary bar is aligned along the line $y$=$x$, or in the \citet{sdeb07} convention, $\Delta\phi$=45$\degree$), then slices at $\pm$45$\degree$ are oriented along the primary bar minor/major axis.  In this case, even in the presence of the secondary bar $<$$v$$>$ and $<$$x$$>$ for the inner primary-bar crossing slices fluctuate noisily about zero.  

This behavior is more than enough to prevent accurate TW estimates for the primary bar (which \citet{rw04} find are less reliable when bars are aligned within $\pm$20$\degree$ of a principle axis), and so it is perhaps not surprising that our TWR estimates at $\pm$45$\degree$ SAs show rather large inaccuracies.  However, what we find requires slightly more interpretation: it is the secondary bar estimate, and not $\Omega_{pb}$, that is compromised.  

Unlike the results in $\S$ \ref{sec:SIM1} with which we could argue that the regularized TWR method is, in principle, fairly equipped to accommodate any alignment of patterns (and where even slices at $\pm$15$\degree$ from the bar axis return accurate TWR estimates), here we are not only dealing with more model parameters (and their intrinsic covariance), but the innermost pattern now has a much smaller extent than the structures in the disk.  Consequently, the secondary bar occupies the smallest fraction of bins in the innermost slices where, moreover, the errors $\sigma_i$ are the largest.  Presumably, in this exploitable state the secondary bar estimate is sacrificed for the primary bar $\Omega_{pb}$ in the regularized calculation; at SA=$\pm$45$\degree$, our measurement of $\Omega_{sb}$ is overestimated by $\sim$25\%, while $\Omega_{pb}$ is good to within 11\%, on average.  (We can assume that such imprecision also reflects the effect of the secondary bar's non-rigid rotation.)  At other slice orientations where the primary bar is more favorably sampled, the trade-off is less severe (though modestly PA-dependent).  

An impromptu calculation using particles in the bulge only (motivated by the \citet{sdeb07} strategy) seems to confirm the influence of the primary bar in the TWR calculation at SA=45$\degree$.  The error in our lone estimate $\Omega_{sb}$=0.42 in this case is significantly less than when the primary bar is present and also comparable to that in bulge-only estimates at other SAs; for the three SAs in quadrant II tested, we find $\Omega_{sb}$=0.43$\pm$0.01.

Even if the TWR method cannot accurately measure the pattern speed of a pulsating nuclear bar--aside from with global regularized solutions, in particular--it is nevertheless appropriate for use in characterizing and constraining $\Omega_{sb}$, especially in the presence of a primary bar. The comparative worth of TWR solutions can be deduced in light of the distinction between the traditional TW and the TWR methods. Our TWR pattern speed measurement for the primary bar is comparable to that from the traditional TW calculation; using slices covering $\vert y \vert\leq$2.25 (such that the bar contribution is maximal in all slices), the TW method returns $\Omega_{pb,TW}$=0.214$\pm$0.011.  The TWR secondary bar estimate, on the other hand, greatly improves upon the TW estimate $\Omega_{sb,TW}$=0.338$\pm$0.149 available using slices covering $\vert y \vert\leq$0.6. 

Since the nuclear bar appears in only a very small number of slices and its contribution is, moreover, easily overwhelmed in the extended TW integrals, the TW method only modestly recovers $\Omega_{sb}$.  In the TWR calculation, however, the few bins that cover the nuclear bar are supplied with large weights $K_{ij}$ compared with the rest of the information along each slice.  Together with a suitable bin width, this effectively isolates information from throughout the nuclear bar extent from that of all other patterns in the disk.  The quality with which $\Omega_{sb}$ can be constrained then depends on how well the contribution from these other patterns is identified and removed from slices intersecting the nuclear bar, and moreover relies on regularization over both sides of the galaxy to construct a global solution.

Aside from the unavoidable error introduced by the non-rigid rotation of nuclear bars expected from N-body simulations, from our current study it seems likely that similar calculations may be limited according to the degree of resolution and the relative size of 
the secondary bar.  That is, the former dictates the precision with which the secondary-to-primary bar transition can be determined, and the latter sets the leverage supplied by the secondary bar to the $\chi^2$.  Furthermore, given that the measurement of an inner, nuclear bar pattern speed is subsequently largely affected by how well the other patterns present in the disk can be measured, success with the TWR method requires appropriate models for $\Omega_p(r)$ that account for such concerns. 
\section{Summary}
\subsection{\label{sec:caveats}Caveats in Applications to Real Galaxies}
The successes of the TWR method as applied to simulated data should be obtainable with real galaxies, provided
that adequate attention is paid to several considerations.  Since the calculation foremost requires integration along 
slices parallel to the galaxy major axis which reflect information from all patterns of interest in the disk, 
observations must be able to present a number of these.  As governed by the resolution or sampling of the 
data, the placement of these slices must be able to define a quadrature wherein the position $y$ of the outermost slice 
corresponds to that 
of the last matrix element along the slice requiring the largest limit of integration $X_{max}$ for integral 
convergence; when convergence can be reached only clearly at the map boundary (perhaps far beyond the 
extent of the patterns), this is essential for correctly accounting for all information along each slice. 

Establishing the slice positions and orientation for an accurate quadrature also clearly requires superior 
knowledge of kinematic and morphological parameters.  In general, the same restrictions to the quality of input data in 
traditional TW calculations apply to the TWR method.  As described in $\S$ \ref{sec:Err}, errors in the assumed PA 
which impair TW estimates can also introduce considerable inaccuracy to the TWR solutions.  The inclination angle of 
the disk must also be well-determined and should be 
preferably restricted to only moderate values (which, beyond the observational requirements of the TW method, will keep 
errors from corrupting the association of information into accurate radial bins).

Additionally, though more easily overcome in traditional TW estimates \citep{mk95}, errors in the systemic 
velocity and galaxy center might prove critical for the TWR solutions since each side of the disk provides an independent 
solution for $\Omega_p(r)$.  Care must be taken not to impair the prevailing symmetry along each slice, given that each 
integral bears more than one estimate.  The incorrect placement of radial bins according to a mis-assigned galaxy center position, for example, could significantly over- or 
under-estimate the actual pattern speed, and moreover make assessment of the true radial variation unlikely.

The regularization procedure developed here itself makes further demands on the quality and amount of information necessary to perform 
the calculation.  But by keeping the amount of information beyond that required of the TW method to a minimum, and using 
standard diagnostics such as Fourier decomposition, the requisite set of a priori assumptions can be invoked quite 
reliably.  As long the information is accessible, 
requiring at the least theoretical motivation to develop testable models, and limited 
in principle only by the quality of information from which it is to be gathered, then the regularization should proceed 
without impediment.

Of course, unlike the simulated galaxies studied here where there is plenty of signal throughout the disk, observations of real galaxies may 
present sensitivity issues.  While regions of low signal-to-noise in the outer disk can be superseded using the `cut' procedure developed here, high quality information from the rest of the disk is an obvious priority for the method; the departures from axisymmetry induced by all patterns present in 
the disk must be clearly detectable.  Not only does the calculation depend on the presence (or lack) of these signatures--in both the surface density and in the 
velocity field--but the mere identification of the number and domain of patterns is critical for developing 
appropriate models for $\Omega_p(r)$.

This latter necessity may be hard met since, for instance, it will be rare to 
observe galaxies with surface densities that can be Fourier-decomposed as cleanly as is 
possible with simulated galaxies.  Furthermore, unlike simulations, it is impossible to
establish whether or not there exists more than a single pattern speed at each measured radius in real galaxies.  
Since the 
models developed with regularization here are incompatible with non-unique pattern speeds, for real galaxies, the choice of when and where to consider a transition or to keep solutions unregularized may be 
based on a more qualitative assessment of where clear structure ends.

Inevitably, the combination of the above considerations (related, overall, to the quality of the data) will
determine the extent to which the model for the true radial behavior of 
$\Omega_p(r)$ can be differentiated from other models.  That is, the $\chi^2$ criterion with which we judge the goodness of 
solutions becomes less discriminating the larger the measurement error $\sigma^{<v>}$.  Since the adopted measurement error $\sigma^{<v>}$ for each slice used in the calculation and in the $\chi^2$ estimator must necessarily incorporate observational errors based on random noise in the data, with severe enough errors different model solutions from real, imperfect data may be indistinguishable.

In addition, systematic errors (likely dominated by PA uncertainty) will undeniably challenge the accuracy of solutions.  In all applications of the 
method it is critical to assess the influence of these errors through direct tests of the sensitivity of solutions to departures from the nominal values of PA, inclination, and kinematic center, for instance. Clearly, this makes 2-D coverage desirable; here, 2-3$\degree$ uncertainties in the PA alone are shown to introduce around 15\% error in measurements of $\Omega_p(r)$ for the barred spiral simulation. 

Insufficient resolution or sampling may also impair TWR solutions from real galaxies.  
A large adopted bin width not only limits the detectable radial variation in 
$\Omega_p(r)$, but also restricts how well multiple patterns can be separated in the resultant quadrature; a mismatch between the actual transition and that to which the solution is confined can have consequences for the estimates of both inner and 
outer patterns.  Naturally, depending on the models to be tested and relative size of the disk, a resolution-constrained bin width is not 
guaranteed to impair solutions for all galaxies. We nonetheless foresee that the only true way to preserve the 
integrity of solutions is with high-resolution observations.

Data cubes lend themselves well to analysis with the TWR method, since unlike long-slit spectroscopic 
observations, the galaxy PA, inclination, systemic velocity and (kinematic) center can be derived with errors from the 
data using a tilted-ring analysis on the first moment of the cube.  Additionally, multiple slices can be defined with 
a single observation.  So between radio and sub-mm investigations of spiral structure, for instance, and IFU spectroscopy with which double bar systems (and eventually double barred spirals, given larger, more sensitive IFUs) can be studied, applications of the TWR method could be extensive.

Of course, like all applications 
of the TW method, the observed tracer must be found to obey continuity and the relation between 
the intensity of the tracer and the surface density must be linear or well-determined everywhere.  Reviews of several possible tracers argued to suitably obey the TW continuity requirement can be found throughout the literature, but we note here that the work of \citet{GD} studying the effect of dust on TW measurements of bars may find meaningful extension in future TWR studies of multiple patterns in late-type galaxies. There, model dust lane features associated with bars introduce errors on the order of 20-40\% \citep{GD}.  In addition to these errors, TWR solutions could possibly be prone to increased error from spiral dust lanes at larger radii.  Though it is beyond the scope of this work to make a detailed assessment of the sensitivity of TWR solutions to dust, we argue that such noise could well be mitigated through the use of regularization, and expect no greater an effect in TWR measurements than TW, which moreover, will be apparent with the use of only optical tracers. 

In the immediate future, we plan to apply the TWR method to several high-resolution
BIMA SONG CO observations of molecule dominated galaxies to search for spiral winding, relations between bar and spiral pattern speeds, and spiral-spiral mode coupling.  (These observations include single-dish data and therefore do not suffer from missing flux which would be a violation of the continuity requirement.)  For those galaxies with ISMs not dominated by molecular gas, we plan to combine the CO with HI data to make total column density maps (assuming the ionized component is negligible).  Since the CO-H$_2$ conversion factor is critical in combining the CO and HI maps (and, of course, in establishing molecule dominance in the former case), it will be necessary to test the sensitivity of the TWR method to the adopted conversion factor for such combinations.

Additionally, since warped disks (common in HI) are a violation of the TW assumptions, we will also perform tests to determine if our cut-off scheme can be used to circumvent the warp and thereby extract solutions from the rest of the disk. For these applications of the TWR method, we plan to construct measurement errors $\sigma$ for each slice that reflect uncertainties related to the flux cutoff chosen in creating the moment maps.  The effect of PA and other systematic errors will be assessed by testing the sensitivity of solutions to departures from the nominal values.
\subsection{\label{sec:prescrip}Applying the TWR Method}
Although in the interest of testing our strategy for each simulated galaxy is somewhat tailored to its unique properties, with the above caveats in mind our studies have enabled us to develop a general and reasonably objective prescription for applying the regularized TWR calculation:

\begin{enumerate}
\item Establish the bin width and the corresponding number of slices (not necessarily uniformly spaced) that are required to achieve converged 
integrals using an $N\times N$ quadrature.  For the purposes of measuring multiple patterns in a single disk, this will
likely extend to the map boundary.
\item Compile a priori information by inspecting the surface density, its Fourier decomposition, and the velocity field for indications of 
patterns and to establish the expected number and domain of measurable pattern speeds.  This should include the identification of regions in the disk 
susceptible to regularization-induced bias.
\item Develop theoretically and/or observationally motivated models which parameterize $\Omega_p(r)$ 
according to the a priori information.
\item Incorporate measurement errors into a single $\sigma^{<v>}$ (and $\sigma$) for each slice.  These should represent uncertainties in the adopted intensity noise level, and/or other random noise-related errors; systematic errors are preferably determined through direct testing (see item 9). 
\item Develop the weighting scheme for a 
reduced $\chi^2$ estimator which accounts for the total degrees of freedom for the models to be tested.  This should 
reflect expectations for which slices, if any, are most critically to be reproduced by the models. 
\item Generate preliminary solutions for the models.  At this point, the degree of regularization required to return 
solutions according to type should be explored.
\item With finalized solutions, use equation (\ref{eq:twr}) to generate a complete set of $<$$v$$>$ for each and calculate the 
corresponding $\chi^2_{\nu}$.  
\item Use the $\chi^2_{\nu}$ to identify the best-fit solution.
\item Test the sensitivity of the results to other systematic effects peculiar to the observation, e.g. adopted PA and/or the 
CO-H$_2$ conversion factor, for instance.
\end{enumerate}

\section{Conclusion}
In this paper we have shown that regularizing the TWR calculation is an effective means of smoothing intrinsically 
noisy solutions for more precise measurement of $\Omega_p(r)$.  Specifically, (barring a large, limiting 
resolution) regularization admits the use of a much smaller bin width than that required to achieve comparable smoothness in the unregularized calculation.  This 
affords improved assessment of radial variation as well as more accurate determination of the 
transitions between multiple pattern speeds (and thus of the values of the pattern speeds themselves, in principle).  Moreover, 
with the regularized TWR calculation, 
different theoretically and observationally motivated models for the radial dependence of $\Omega_p(r)$ can be 
tested in fairly short time and with only the minor addition of information compared to the unregularized TWR and TW methods.

With a simple scheme for generating $n^{th}$ order polynomial solutions which can be incorporated into step 
models, we have shown that the TWR method can be used to parameterize the radial domains of multiple pattern speeds.  
Together with a priori information identifying zones in the disk which may be incompatible with measurement (either 
because they are characterized by low signal-to-noise or show no evidence for a pattern), we can further constrain 
the extent of patterns while optimally reducing regularization-induced bias in pattern speed solutions.

As applied to three simulated galaxies, we find that the 
TWR method developed in this manner
performs with a high degree of accuracy (with less than 15\% error) both in measurement and in extracting information 
about the true functional form for the pattern speed.  Tests on a simulation of a barred spiral galaxy indicate that 
not only can the constant pattern speed for a relatively weak spiral be reliably reproduced, but information about both 
the pattern speed and the radial extent of the bar pattern can also be extracted.  (Indeed, we find that the bar pattern 
speed estimate is strengthened by the proper use of information from beyond the bar end.)  And though the bar pattern speed
estimates is highly susceptible to systematic errors--with PA errors introducing the largest uncertainty to TWR pattern speeds, as with TW estimates--we find that the identification of the transition between the two is relatively stable.

The TWR method can also be effectively employed to measure patterns that are winding in nature. In a simulation 
of a two-armed spiral, the best-fit TWR solutions from several slice orientations correctly reproduce the high-order 
radial variation of the pattern speed, despite modest indication that not all orientations supply the same authority 
(this, of course, would seem to depend on the morphology of this spiral, in particular).  Indeed, though the TWR method 
can, in principle, handle any (presumably random) alignment of patterns, in all of the simulations studied, slice 
orientations which provide the most uniform coverage of the patterns are preferred.  This is of particular 
importance for nuclear bar pattern speed measurement, as found in tests of the method 
on a double barred simulation.  Since the innermost bins which provide the foremost leverage on the nuclear bar are also the most susceptible to errors from throughout the disk, confident measurement requires all other patterns to be well constrained.

In principle, comparable accuracy should be achievable on real galaxies.  However, these tests do not constrain 
how well the TWR method can perform under severe observational limitations which may commonly arise.  Not only can determinations of the PA, 
inclination, and dynamical center be subject to considerable errors given low-quality data, but identifying constraints 
on the patterns present in the disk to be incorporated into models for $\Omega_p(r)$ could prove challenging.  
Additionally, though regularization can reduce the impact of noise on solutions, large measurement errors for each slice could make discriminating between several possible models for $\Omega_p(r)$ difficult.  And most critically, since the nature of the numerical 
calculation relies on a relatively small bin width to achieve its greatest accuracy, without high resolution, some observations may not afford practical solutions.

Nevertheless, if restricted to high resolution observations with adequate sensitivity, and given radially stable 
kinematic parameters, TWR solutions can be used to study the connection 
between multiple patterns and the nature of spiral winding.  So, too, can we expect progressively more satisfactory applications of the method; though the number of galaxies to which the method can be successfully applied is limited by the current generation of instruments, in the future, larger IFUs, ALMA, and eventually, SKA should yield much higher quality data with larger areal coverage and higher angular resolution.  This prospect in itself should warrant future studies with the TWR method.  \\

This research was supported in part by NASA through the American Astronomical Society's Small Research Grant Program.  S. E. M. acknowledges support from a NASA-funded New Mexico Space Grant Consortium Graduate Research Fellowship.  V. P. D. is supported by an RCUK Fellowship at the University of Central Lancashire.  J. S. acknowledges support from a Harlan J. Smith fellowship.  This material is based on work partially supported by the National Science Foundation under grant AST 03-06958 to R. J. R.  
%
% References
%
%\bibliography{}

% LocalWords:  Meidt Merrifield NG Debattista Juntai Shen Pertti Rautiainen TWR
% LocalWords:  Oulu ss syg SPH TW unwarped axisymmetric dgs NGC dw dca cda MRM
% LocalWords:  inner's BIMA discretized discretization arcsin dr dx Volterra mk
% LocalWords:  ij zrm NN backsubstitution kk kpc Tikhonov tich rrrrrrrrrrr x's
% LocalWords:  minima im unregularized CCW kinematical PAs rms BStw PAerror tw
% LocalWords:  misassigning nonaxisymmetry minorly twslice PA's fourier Stw ph
% LocalWords:  Ssol Ssolb unrigid Prieto MNRAS Corsini astro Sellwood Sevenster
% LocalWords:  Carollow Wadsley Kuijken Kormendy Pohlen Maciejewski Masset Salo
% LocalWords:  Wallin Sparke Sygnet Tichonov Arsenin Hamabe ApJS Zimmer ds xy
% LocalWords:  binnings corotation debPA BSsol SIMI di SIMIII Btw

\end{document}